# Fourier series-based modelling of the effects of thermal coupling on the transient dynamics of component loops in a Coupled Natural Circulation Loop

Prasanth Subramaniyan, Akhil Dass, Shivangi Tiwari and Sateesh Gedupudi[1]

*Heat Transfer and Thermal Power Laboratory, Department of Mechanical Engineering, IIT Madras, Chennai 600036, India*

## Abstract

Energy efficiency in process industry and passive safety systems in nuclear power plants necessitate the use of buoyancy driven heat exchangers. The current study presents a 1-D numerical model of the buoyancy driven fluid system in a Coupled Natural Circulation Loop (CNCL) comprising of water as the working fluid. In this study water at different temperatures is considered as the operating fluid in each of the loops. The study employs a 1-D model derived using a semi-analytical Fourier series. A validation study was carried out using the relevant literature to verify the mathematical model. A detailed parametric study on individual NCLs and their coupled system was conducted by varying the cross-sectional area keeping the other parameters constant to gain insights into the effect of thermal coupling on the long-term transient dynamics of each of the individual loops of the CNCL system, for stable, neutrally stable, and unstable conditions. The dynamics were analysed using the difference between transient buoyancy and viscous forces, and it was found that the overall heat transfer coefficient influences the coupling behaviour and the dynamics of the component loops in a CNCL. At lower values of the overall heat transfer coefficient, the component loops in the CNCL nearly retain their independent behaviour, i.e., the component loops hardly influence each other. It was also found that the temperature dependent fluid properties influence the stability of the CNCL system in some cases.

*Keywords:* Coupled Natural Circulation Loop, Fourier series, 1-D model, buoyancy, transient dynamics.

## 1. Introduction

A Natural Circulation Loop (NCL) works on the principle of natural circulation, a buoyancy-driven phenomenon occurring in a closed loop present in a body force field with external thermal stimulation. A Coupled Natural Circulation Loop (CNCL) comprises two or more NCLs that are thermally coupled, facilitating heat transfer between them by means of a common heat exchanger. Depending on the heat exchanger orientation, the CNCL system can be further classified into horizontal (HCNCL) or vertical (VCNCL) systems. Buoyancy driven heat exchangers eliminate the need for pumps (and hence pumping power), offer enhanced safety and reliability and have simple fabrication process and smooth operation with no moving parts. CNCLs are used as passive residual heat removal systems in case of pump failures in nuclear reactors.

[1]Corresponding author. Tel.: +91 44 2257 4721, Email: sateeshg@iitm.ac.in

## 1.1. Behaviour of NCL systems

Keller (1966) was the first to explain that the dynamics of an NCL depended on the buoyancy and viscous forces alone. The chaotic behaviour of an NCL system was observed by Welander (1967) and explained the cause due to thermal anomalies, later it was experimentally investigated by Creveling et al. (1975). The steady state and transient dynamics of pressure difference in a toroidal NCL was presented by Mertol et al. (1984). An experimental investigation on the effects of inclination and the effects of changing heat input on an NCL was reported by Acosta et al. (1987), and demonstrated the existence of multiple steady-state solutions. Vijayan et al. (1991) reported a comparison of loss coefficients and Nusselt numbers (Nu), a generalized loss coefficient was derived from Grashof number (Gr). Experimental investigation of chaotic oscillations and the effect of axial conduction and inclination on the transient dynamics of a toroidal loop was reported by Jiang et al. (2002). An experimental study to represent the correlation between Reynolds number (Re) and Grashof number (Gr) was reported by Vijayan (2002). Garibaldi and Misale (2008) investigated single-phase mini loop NCLs with different heights. The report analysed transient and steady state behaviour of water and FC43 at different inclinations and power input. Vijayan et al. (2008) studied the effect of loop diameter for both single-phase and two-phase natural circulation loops. The study revealed that an increase in the diameter of the single phase NCL system led to increase in instabilities, whereas, the stability of two-phase loops is found to enhance with increase in loop diameter. An experimental examination conducted by Misale et al. (2011) showed the influence of thermal boundary conditions on the dynamic behaviour of a single-phase NCL. The study presented the implementation of Ultrasound Pulsed Doppler Velocimetry (UPDV) for the first time. An experimental analysis was carried out by Swapnalee and Vijayan (2011) to obtain the friction factors in the laminar, transition and turbulent regimes of an NCL system. The determined friction factors were employed to obtain the domains of stability for all flow regimes. Ma et al. (2007) studied the Lead Bismuth Eutectic (LBE) NCL system experimentally, and conducted a stability analysis for various input power. The start-up transience and effect of different parameters on natural circulation flow rate was presented. Kang et al. (2013) investigated the dynamics of NCL with liquid gallium as working fluid and an experimental study was conducted to determine the Nusselt number at the cooling section. Cheng et al. (2019) conducted experimental investigation of the thermal-hydraulic performance of natural circulation in a mini-loop and presented a correlation for Nusselt number and Rayleigh number.

## 1.2. Modelling of NCL systems

Zwrin and Greif (1979) were the first to develop a 1-D model based on approximate equations for the non-linear system and obtain transient solutions. Stability analysis was conducted by Hart (1984) suggesting a Fourier series based 1-D model to convert the system to a set of Ordinary differential equations (ODEs). A five mode Bessel-function based model was presented by Yorke J and Yorke E (1987) to convert the PDEs to ODEs, and explained that a function with 4-5 modes is sufficient to capture the dynamics of the system. The effect of loop diameter and Stanton number (St) on stability behaviour was presented by Vijayan and Austregesilo (1994) using both experimental and 1-D model. A simulation using ATHLET code to predict the amplitude and phase of fluid temperature oscillations was given by Vijayan et al. (1995). A 2-D CFD study to analyse the effect of wall conduction and pipe thermal capacity was first reported by Misale et al. (2000). The deviations observed with the respect to 1-D predictions was accredited to the axial conduction effects. The effect of gravity on the transient and stability behaviour of a rectangular NCL was presented by Cammarata et al. (2004) using an experimental and 1-D model. The value of gravitational field that separates the region of stable dynamics and chaotic behaviour was determined. A detailed transient study of the rectangular NCL using FDM, linear analysis using perturbation and RELAP5 system code was conducted by Mousavian et al. (2004). A finite element (FE) simulation for different temperature input conditions on the steady-state and temporal behaviour of rectangular NCL was conducted by Rao et al. (2005b). A temporal dynamic pressure distribution along with the temporal variation in temperature and development of flow rate was reported by Rao et al. (2005c). Basu et al. (2007) employed a 1-D model using finite volume method (FVM) to study the effect of ambient temperature and obtained the steady state behaviour of a rectangular NCL. It was reported that for the same ambient temperature, the effect of heat loss is higher for a smaller diameter loop due to the reduced heat transfer surface. A non-dimensional 1-D modelling of an NCL coupled to a thermal storage was carried out by Benne and Homan (2008). The non-dimensional numbers accounted for parameters such as aspect ratio of the tank, the rate of heat input, as well as the relative length and diameter of the portion of the loop dominating the frictional loss. Lin et al. (2008) incorporated conjugate heat transfer characteristics in the NCL system. A 1-D finite difference method (FDM) model was developed to investigate the effect of wall conduction effect. One of the first 3-D CFD studies for a single-phase NCL at steady state was simulated by Angelo et al. (2012). They reported the influence of vortex and swirl at the two ends of heater section. Basu et al. (2012) carried out a 1-D numerical investigation of NCLs and identified that the neutral stability condition corresponds to a larger power supply, with coolant condition remaining the same. Their study also reported that the toroidal loop was observed to be more stable compared to an identical rectangular loop subjected to bottom heating and top cooling, due to the smaller magnitude of developed buoyancy force. A CFD analysis of single-phase carbon dioxide ($CO_2$) was conducted by Yadav et al. (2012). The study presented a correlation between the Reynolds number and the Grashof number and concluded that to achieve the maximum heat transfer rate, operating condition should be chosen near the pseudo-critical region in supercritical zone. The simulations were able to capture the velocity fluctuation at the heat exchanger due to cross sectional area variation and the vortex and swirl effects before and after the heater location. Basu et al. (2013a) developed a unified model for explaining steady-state operation of an NCL, and concluded that the Richardson number (Ri) as characterizing parameter is a better choice compared to the conventional Grashof number (Gr). In a CFD study of a 2-D thermosyphon, Louisos et al. (2013) used non-dimensional parameters that captured the physics of the system, and explained that four distinct flow regimes exist. A detailed study on the influence of geometry and operating parameters was conducted by Basu et al. (2013c). The work accounted for different geometric dimensions including loop diameter, height, width, resistance, and inclination. It was reported that a smaller diameter loop was found to be more stable due to enhanced friction. Stability behaviour was also enhanced with a longer heating section and wider loop, and significantly improved with increase in height. Incorporating minor frictional resistances in the form of valves or orifices suppressed flow instability. Inclination in the form of tilts in the range of 10-15 degrees was sufficient to suppress unstable fluctuations even at large power levels. A 1-D MATLAB based Simulink model was developed by Lu et al. (2014) to conduct transient analysis of LBE as working fluid in an NCL. The model incorporated the influence of heat loss and wall thickness effect. Numerical simulations were conducted for increasing Rayleigh numbers to investigate various flow dynamics. The effects of internal heat generation on the stability of NCL systems was investigated by Ruiz et al. (2015) using a linear semi-analytical method and numerical non-linear method. Goudarzi and Talebi (2015) conducted a non-linear stability analysis and investigated the entropy generation in the loop. Parametric study involving loop diameter, heater and cooler length, and effect of orifices was presented. Srivastava et al. (2016) conducted transient studies of an NCL system including startup of natural circulation, loss of heat sink, heater trip and step change in power, with molten nitrate salt as working fluid. Their study used a 1-D LBENC code with experimental validation. An experimental and 3-D CFD study for steady-state and transient characteristics of a molten salt NCL was reported by Kudariyawar et al. (2016). A linear and non-linear stability analysis on volumetric heat generation on the increase of stability domain was carried out by Pini et al. (2016) for different heater and cooler configurations. Cammi et al. (2016), using a CFD and 1-D RELAP model concluded that the inclusion of wall thickness effect leads to an increase in the domain of stable operation of the NCL system. A 3-D CFD study was conducted by Krishnani and Basu (2016) to identify the limitations of the Boussinesq approximation. It was concluded that the Boussinesq model can be employed to get a rough estimate at lower powers, provided the fluid temperature variation is limited to 10° C and average fluid temperature is around the reference temperature. A 3D computational and 1-D model developed by

Krishnani and Basu (2017) conclude that the loop inclination has a strong stabilizing influence. It was also reported that the loop experiences unstable oscillations with increase in both heater power and sink temperature. Wu et al. (2017) proposed a new correlation to obtain the mass flow rate of natural circulation in a complex system in terms of the heating power and ratio of local pressure loss coefficient and friction coefficient. The emphasise on the necessity to consider the variation of heat transfer coefficient due to the velocity field perturbations was presented by Luzzi et al. (2017) with the help of a 3-D CFD simulation and 1-D model. Later, this was validated by Cammi et al. (2017) by utilising the information entropy method and obtaining the stability maps of the NCL systems. Cheng et al. (2018) employed a FDM to discretize the energy and momentum equations for obtaining the transient behaviour of the single-phase NCL and presented the entropy generation analysis. Stability maps were generated based on the Nyquist criterion. Saha et al. (2018) conducted a dynamic characterization of a single-phase square NCL. It was observed that for a stable oscillatory flow increasing heater power causes increased amplitude for mass flux or temperature difference, and further increase in heater power can cause flow reversal (chaotic). It was shown that the Fast Fourier Transform (FFT) of the oscillatory region phase plots have only one fundamental frequency and other frequencies are the harmonics. In the flow reversal region, phase plots show Lorenz like strange attractor and no fundamental frequency can be identified in the FFT. Nadella et al. (2018) investigated the effects of wall thermal inertia, local pressure losses and variable heat exchange on the stability behaviour of an NCL system. In a 3-D CFD study, Zhu et al. (2019) suggested that modifying the NCL shape from rectangular to a mobius strip reduced the flow oscillation without increasing the pressure loss. It was reported that the oscillations decreased with enclosed area, angular acceleration, and the angle between them. The startup dynamics of a Passive residual heat removal system (PRHRS) after incorporating fully developed friction and heat transfer assumptions was reported by Zhao et al. (2019), which incorporated a 1-D model and experimental setup. Flow reversal predictions in a chaotic system using symbolic time series analysis was conducted by Saha et al. (2020), considering wall conduction effects. Goyal et al. (2020) studied the chaotic nature of NCL system predicted via linear stability analysis and concluded that the system may diverge for slightly large magnitude of perturbations for the regions identified as stable through linear stability analysis. The model emphasises the inadequacy of the linear analysis to comprehend the dynamics of such a complex system.

## 1.3. CNCL systems

The concept of multi-loop thermosyphons was first studied by Sen and Fernández (1985), who presented a simplified 1-D analytical model excluding the axial conduction effects. Davis and Roppo (1987) developed two Lorenz oscillators that were linearly coupled to conduct a stability analysis, but there was no validation provided for their model and the explanation of transient dynamics was missing. Later, their work was validated by Erhard (1988). Salazar et al. (2012) were the first to perform steady state analysis and demonstrated the possibility of multiple steady-state solutions using a rectangular CNCL. Additional simplifications in the frictional forces and no axial conduction effect were considered. A transient study of the CNCL performance employing water as the test fluid was carried out by Ghoneimy and Dougall (2017). Experiments were conducted to characterize the dynamic system behaviour and to study the capability of the CNCL as a safety mechanism for decay heat removal. The stability of coupled loops at different input powers was studied by Zhang et al. (2015). The steady-state 1-D equations incorporating the axial conduction and conjugate effect of multi-loop CNCLs at different configurations (series and parallel) was presented by Vijayan et al. (2019). A 1-D single phase semi-analytical modelling of the transient CNCL system having flat plate heat exchanger formed via the coupling of constituent square NCLs was presented by Dass and Gedupudi (2019). Non-dimensional numbers were employed to study the transient dynamics of the CNCL system at different heater and cooler configurations. Another study conducted by Dass and Gedupudi (2021a) incorporated the effects of wall conduction in the heat exchanger section and the inclination on the system. The effect of inclination on the buoyancy driven system indicating the occurrence of the heat transfer coefficient jump and the hysteresis in tilted NCLs and CNCLs was reported by Dass and Gedupudi (2021b). A linear stability analysis of the CNCL system employing a Fourier series based 1-D model was carried out and the stability maps were developed by Dass and Gedupudi (2022). It was observed that the long-term transient behaviour of the CNCL system predicted by a stability map is independent of the initial condition. Adinarayana et al. (2022) investigated the transient performance of series-coupled NCLs using a Finite Volume numerical model. The model incorporated multiple features including equation of state, wall conduction and different heater-cooler configurations. An experimental study to analyse the stability behaviour of series coupled rectangular loops was reported by Elton et al. (2022). The study confirmed the hysteresis phenomenon and revealed the transmission of instability from one loop to another, with attenuated oscillations attributed to the wall damping effect. Another study conducted by Adinarayana et al. (2023) using two-phase models predicted the start-up transients and analysed the influence of heater-cooler orientations and system conditions in a coupled system. A parametric study based on different geometrical and operating conditions was reported. A station blackout scenario (SBO) was simulated using a Natural Circulation Mechanism Experimental Setup (NCMES) by Chang et al. (2023). The calculated results showed a good agreement with experimental data and concluded that under an SBO scenario a triple-loop coupled system is sufficient to remove the residual and sensible heat of the fluid. Benzoni et al. (2023) presented a 1-D model to study the effect of numerical integration algorithm on the simulation behaviour of the coupled DYNASTY-eDYNASTY natural circulation loops. The influence of eDYNASTY loop on the primary (DYNASTY) loop was reported along with a parametric analysis of the system's behaviour due to their cooler configuration.

From the literature review it can be concluded that the studies dedicated to the influence of thermal coupling on the long-term transient behaviour of component NCLs of a CNCL appear to be scarce. Most of the literature is based on the use of non-dimensional parameters to explain the transient dynamics and the incorporation of temperature dependent fluid properties was not considered for 1-D CNCL models available in the literature. The objective of the present study is to investigate the effect of thermal coupling on the long-term transient dynamics of component NCLs of a CNCL system by comparing their behaviour with the behaviour of independent NCLs under similar initial and operating conditions, taking into account the temperature dependent fluid properties. The study is also aimed at gaining physical insight into the coupling effect and the system dynamics by studying the difference between the transient buoyancy and viscous forces per unit volume.

**Nomenclature**

| | |
|---|---|
| $a$ | Thermal diffusivity (m$^2$/s) |
| $A_{cs}$ | Flow cross-section area (m$^2$) |
| $C_{pi}$ | Specific heat capacity (J/kg-K) |
| $D_h$ | Hydraulic diameter (m) |
| $f_{Bi}$ | Buoyancy force per unit volume (N/m$^3$) |
| $f_{Vi}$ | Viscous force per unit volume (N/m$^3$) |
| $g$ | Gravitational constant (m/s$^2$) |
| $h_1$ | Heat transfer coefficient of fluid 1 (W/m$^2$-K) |
| $h_2$ | Heat transfer coefficient of fluid 2 (W/m$^2$-K) |
| $L$ | Loop height (m) |
| $L_1$ | Loop width (Loop 1) (m) |
| $L_2$ | Loop width (Loop 2) (m) |
| $L_h$ | Heated length ($L_i - D_h$) (m) |
| $n$ | Number of bends on the component loops |
| $P_{0,i}$ | Initial pressure (atm) |
| $P_{cs}$ | Wetted perimeter of the cross-section (m) |

| | | |
|---|---|---|
| $Q$ | Heat flux (W/m$^2$) | |
| $Q_H$ | Heat input at the heater section (W) | |
| $Q_{HX}$ | Heat transfer at the heat exchanger section (W) | |
| $r$ | Radius of curvature of the bend (m) | |
| $T_{0,i}$ | Initial temperature (K) | |
| $T_{1,avg}$ | Average temperature of loop 1 (K) | |
| $T_{2,avg}$ | Average temperature of loop 2 (K) | |
| $T_1$ | Temperature of loop 1 (K) | |
| $T_2$ | Temperature of loop 1 (K) | |
| $t$ | Time (s) | |
| $u_{0,i}$ | Initial velocity (m/s) | |
| $u_i$ | Velocity of fluid (component loops) (m/s) | |
| $u'_i$ | Velocity of fluid (independent NCL) (m/s) | |
| $U$ | Overall heat transfer coefficient at heat exchanger section (W/m$^2$-K) | |
| $w$ | Width of the cross-section (m) | |
| $d$ | Depth of the cross-section (m) | |
| $x$ | Distance from the origin (m) | |

**Greek letter**

| | |
|---|---|
| $\beta_i$ | Volumetric thermal expansion coefficient (K$^{-1}$) |
| $\kappa$ | Thermal conductivity (W/m-K) |
| $v$ | Kinematic viscosity (m$^2$/s) |
| $\rho_i$ | Density of fluid (kg/m$^3$) |
| $\tau_i$ | Wall shear stress (Pa) |

**Fourier coefficient**

| | |
|---|---|
| $\alpha_n$ | Fourier coefficient of $h'_1(x)$ |
| $\beta_n$ | Fourier coefficient of $h'_2(x)$ |
| $\varepsilon_n$ | Fourier coefficient of $h_1(x)$ |
| $\zeta_n$ | Fourier coefficient of $h_2(x)$ |
| $\gamma_n$ | Fourier coefficient of $T'_1(x, t)$ |
| $\delta_n$ | Fourier coefficient of $T'_2(x, t)$ |
| $\theta_n$ | Fourier coefficient of $T_1(x, t)$ |
| $\varphi_n$ | Fourier coefficient of $T_2(x, t)$ |
| $\chi_n$ | Fourier coefficient of $\omega_1(x)$ |
| $\psi_n$ | Fourier coefficient of $\omega_2(x)$ |
| $A_n$ | Fourier coefficient of $f_1(x)$ |
| $B_n$ | Fourier coefficient of $f_2(x)$ |

**Piece-wise functions**

| | |
|---|---|
| $f_i(x)$ | Function representing the geometry of the loop |
| $H_i(x)$ | Function representing the location and magnitude of the heat flux boundary condition and heat transfer due to convection. |
| $h'_i(x)$ | Function representing heat flux boundary condition of component NCLs |
| $h_i(x)$ | Function representing heat flux boundary condition of coupled NCLs |
| $h_{ci}(x)$ | Function representing heat transfer at heat exchanger section due to convection |
| $\omega_i(x)$ | Function representing thermal coupling |
| $j'_i(x)$ | Function representing $h'_i(x)$ |
| $j_i(x)$ | Function representing $h_i(x)$ |

**Non-dimensional numbers/parameters**

| | |
|---|---|
| $\alpha$ | Cross section aspect ratio ($w/d$) |
| $f_D$ | Darcy friction factor |
| $f_F$ | Fanning friction factor |
| $Gr_m$ | Modified Grashof number |
| $K$ | Bend loss coefficient |
| $N_g$ | Geometric parameter |
| $Pr$ | Prandtl number |
| $Re$ | Reynolds number |
| $Nu$ | Nusselt number |
| $P$ | Constant in $f_F$ (varies in flow regimes) |
| $b$ | Constant in $f_F$ (varies in flow regimes) |

**Subscript**

| | |
|---|---|
| 0 | Initial value |
| $i$ | Indicating Loop 1 ($i = 1$) or Loop 2 ($i = 2$) |
| $n$ | Number of nodes (Fourier coefficient) |
| $avg$ | Average value |

**Abbreviations**

NCL    Natural Circulation Loop

CFD    Computational Fluid Dynamics

CNCL  Coupled Natural Circulation Loop

HHVC  Horizontal heater vertical cooler

VHHC  Vertical heater horizontal cooler

SSPRK Strong Stability Preserving Runge-Kutta

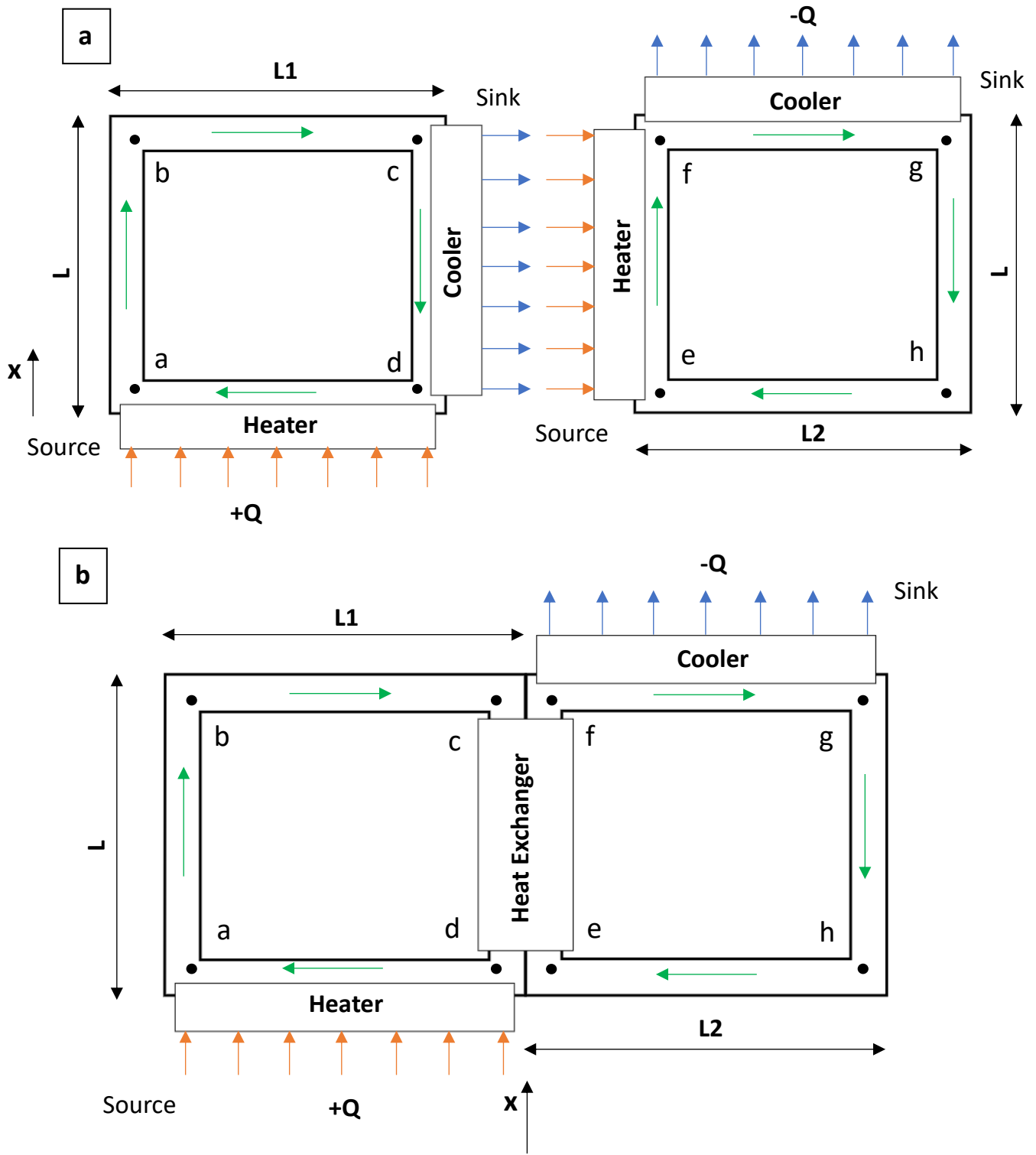

Fig.1. Schematic representation of (a) Independent NCLs loop 1 and loop 2 (b) Coupled NCL system.

## 2. Methodology and Modelling

*2.1. Geometry*

In this study a CNCL system comprising two component NCLs of HHVC (Loop 1) and VHHC (Loop 2) configuration coupled using a common flat plate heat exchanger is considered for the analysis of long-term transient behaviour. The heat exchanger acts as a heat sink for the first loop and as a heat source for the second loop. The limb consisting of the common heat exchanger has a dimension *L*, and the lengths of loop 1 and loop 2 are taken as *L1* and *L2*, respectively. Both the loops have a rectangular cross section which has the same flow cross-sectional area due to the constraints imposed at the heat exchanger section. The width and depth of the cross-section are denoted by *w* and *d*, respectively.

The hydraulic diameter is evaluated as four times the flow cross-sectional area divided by wetted perimeter.

$$D_h = \frac{4A_{cs}}{P_{cs}} = \frac{2wd}{(w+d)} \quad (1)$$

*2.2. Governing equations and Correlations*

The system entirely depends on the buoyancy forces for the propagation (circulation) of the fluids in the loops as stated by Keller (1966). The Boussinesq approximation was employed based on the conclusion drawn by Krishnani and Basu (2016). 1-D momentum and energy conservation equations, incorporating the fluid axial conduction effects presented by Dass and Gedupudi (2019) (based on force and energy balance on an infinitesimal element) is considered for mathematical modelling. A Fourier series based 1-D model is formulated with water as the working fluid to represent the dynamical CNCL system and its component loops. The model was derived from the insights presented by Hart (1984) and Rodríguez-Bernal and Van Vleck (1998).

The momentum and energy equations are represented as:

$$\rho_i \frac{du_i(t)}{dt} + \frac{P_{cs}\tau_i}{A_{cs}} = \frac{\rho_i g \beta_i}{2(L+L_i)} (\oint (T_i - T_{0,i}) f_i(x) dx) - \frac{nK\rho_i u_i^2}{4(L+L_i)} \quad \left(i = \begin{cases} 1, for\ loop\ 1 \\ 2, for\ loop\ 2 \end{cases}\right) \quad (2)$$

$$\frac{\partial T_i}{\partial t} + u_i(t)\frac{\partial T_i}{\partial x} = H_i(x,t) + a_i \frac{\partial^2 T_i}{\partial x^2} \quad (3)$$

The wall shear stress term in the momentum equation depends on the geometry of the flow cross sectional area, along with the flow velocity and density of the fluid.

$$\tau_i = \frac{\rho_i u_i^2 f_F}{2} \tag{4}$$

The $H_i(x,t)$ in the energy equation indicates the location and magnitude of the heat flux boundary condition and the effect of heat transfer due to convection (for a coupled system) at the heat exchanger section.

For a CNCL system:

$$H_i(x,t) = h_i(x) - h_{ci}(x) \tag{5}$$

For an NCL system:

$$H_i(x,t) = h'_i(x) \tag{6}$$

Where $h_i(x)$ and $h'_i(x)$ denotes the location and magnitude of the heat flux boundary condition of component and individual loops (1 and 2) respectively and $h_{ci}(x)$ denotes the heat transfer at the heat exchanger section due to convection.

$$h_i(x) = \frac{QP_{cs}}{\rho_i C_{pi} A_{cs}} j_i(x) \tag{7}$$

$$h_{ci}(x) = \frac{d}{\rho_i C_{pi} A_{cs}} U(T_i - T_j)\omega_i(x) \tag{8}$$

$$h'_i(x) = \frac{QP_{cs}}{\rho_i C_{pi} A_{cs}} j'_i(x) \tag{9}$$

*2.3. Piecewise functions and system conditions (initial and boundary)*

In the momentum equation the function $f_1(x)$ and $f_2(x)$ represents the geometry of the individual NCLs.

$$f_i(x) = \begin{cases} 1 & 0 < x < L \\ 0 & L < x < L + L_i \\ -1 & L + L_i < x < 2L + L_i \\ 0 & 2L + L_i < x < 2(L + L_i) \end{cases} \tag{10}$$

The function $\omega_i(x)$ in Eq. (8) represents the function that couples the individual NCLs (Loop 1 and Loop 2). The coupling is enabled at the heat exchanger section $(0 < x < L)$.

$$\omega_i(x) = \begin{cases} 1 & 0 < x < L \\ 0 & L < x < L + L_i \\ 0 & L + L_i < x < 2L + L_i \\ 0 & 2L + L_i < x < 2(L + L_i) \end{cases} \quad (11)$$

*2.3.1 Boundary Conditions*

(a) NCL $(H_i(x,t) = h'_i(x))$

A constant heat flux boundary condition is imposed at the heater (+Q) and cooler (-Q) of the respective independent NCLs.

$$\frac{QP_{cs}}{\rho_i C_{pi} A_{cs}} = \frac{2Q(w+d)}{\rho_i C_{pi} wd} \quad (12)$$

$$NCL\ 1: j'_1(x) = \begin{cases} 0 & 0 < x < L \\ 0 & L < x < L + L_1 \\ -1 & L + L_1 < x < 2L + L_1 \\ 1 & 2L + L_1 < x < 2(L + L_1) \end{cases} \quad (13)$$

$$NCL\ 2: j'_2(x) = \begin{cases} 1 & 0 < x < L \\ -1 & L < x < L + L_2 \\ 0 & L + L_2 < x < 2L + L_2 \\ 0 & 2L + L_2 < x < 2(L + L_2) \end{cases} \quad (14)$$

(b) CNCL $(H_i(x,t) = h_i(x) - h_{ci}(x))$

The boundary conditions are imposed such that the loop 1 consisting of the horizontal heater at the bottom is provided a constant heat flux of +Q, and the same amount of heat flux -Q is ejected from loop 2 comprising of a horizontal cooler. Hence, a constant heat flux boundary condition is maintained at the heater and cooler of the respective loops.

$$Loop\ 1: j_1(x) = \begin{cases} 0 & 0 < x < L \\ 0 & L < x < L + L_1 \\ 0 & L + L_1 < x < 2L + L_1 \\ 1 & 2L + L_1 < x < 2(L + L_1) \end{cases} \quad (15)$$

$$\text{Loop 2: } j_2(x) = \begin{cases} 0 & 0 < x < L \\ -1 & L < x < L + L_2 \\ 0 & L + L_2 < x < 2L + L_2 \\ 0 & 2L + L_2 < x < 2(L + L_2) \end{cases} \quad (16)$$

*2.3.2 Initial Conditions*

The same initial conditions are provided for both individual and coupled NCLs for conducting a long-term transient analysis and studying the effect of coupling. These initial conditions are applied to the momentum and energy conservation equations (Eq. 2 and 3) of the NCL and CNCL systems.

$$u_1(t=0) = u_{0,1}, \qquad u_2(t=0) = u_{0,2} \quad (17)$$

$$T_1(x, t=0) = T_{0,1}, \qquad T_2(x, t=0) = T_{0,2} \quad (18)$$

The initial temperatures ($T_{0,i}$) were taken as 313.15 K (loop 1) and 303.15 K (loop 2) for both the individual NCLs and component loops of the coupled system. The pressure ($P_{0,i}$) was kept constant at 1 atm, and a gravitational constant ($g$) of 9.81 m/s² was considered. The flow observed in a CNCL with HHVC-VHHC arrangement always exhibits a counter flow behaviour at the heat exchanger section (Dass and Gedupudi (2019)). Due to this reason the initial velocity considered for loop 1 and loop 2 should be carefully incorporated with appropriate signs for the simulation of individual and component loops in the coupled system. For the present study an initial velocity ($u_{0,i}$) of magnitude 0.001 m/s was considered for individual NCLs (1 and 2). The component loops were initialised with -0.001 m/s (for loop 1) and 0.001 m/s (for loop 2). The reason for initialising with a lower velocity value is to meet the restrictions imposed by the implemented ODE solver. Moreover, it was identified by Dass and Gedupudi (2019) that the initial velocities can only have an impact during the initial transients and will not affect the long-term behaviour irrespective of the nature of oscillations being stable or chaotic (Dass and Gedupudi, 2022).

*2.4. Friction factor correlations*

In an individual or coupled NCL system the buoyancy forces are counteracted by the viscous forces. The system attains a steady-state when these forces are balanced. The viscous forces are a result of wall shear stress, geometry of the cross-section, and bend losses. The friction factors contribute towards the wall shear stress term, and vary based on the flow regimes (laminar, transition or turbulent). Friction factors are predominantly dependent on the geometry of the flow cross-sectional area in addition to the Reynolds number for laminar flow. In transition and turbulent regimes, these factors depend on the surface roughness of the pipes and geometry in addition to the Reynolds number.

The classification of the flow regimes for an NCL system was made following the study presented by Swapnalee and Vijayan (2011), where the flow regimes are categorised as laminar ($Re<800$), transition ($800<Re<3000$), and turbulent ($Re>3000$).

In the present study, the flow cross-section is taken as a rectangular duct and a small surface roughness (hydraulically smooth) is considered. All the friction factors considered are the Fanning friction factors.

$$f_F = \frac{P}{Re^b} \tag{19}$$

*2.4.1 Laminar flow (0<Re<800)*

A pipe with circular cross section and hydraulically smooth surface has the following Fanning ($f_F$) friction factor in the laminar regime:

$$f_F = \frac{16}{Re} \tag{20}$$

In the current study, a cross-sectional aspect ratio ($\alpha = w/d$) of 0.43 was taken for a rectangular duct in all the simulations. The Fanning friction factor for the corresponding '$\alpha$' in the laminar flow regime for non-circular pipes was interpolated from Cengel and Ghajar (2015). The value of $P$ was obtained as 16.183 and $b$ as 1.

*2.4.2 Transition flow (800<Re<3000)*

The empirical relation for transition flow was obtained from the work of Swapnalee and Vijayan (2011).

$$f_F = \frac{0.3015}{Re^{0.416}} \tag{21}$$

*2.4.3 Turbulent flow (Re>3000)*

For a fully developed turbulent flow in rough pipes, the friction factor is a function of the Reynolds number and the pipe relative roughness height ($\epsilon$). In our present study, it was assumed that the surface roughness is very small and the friction factor for turbulent flow in smooth pipes can be approximated empirically by the Blasius equation:

$$f_F = \frac{0.079}{Re^{0.25}} \tag{22}$$

*2.5. Heat transfer coefficient*

The governing equations presented for CNCL depict that an overall heat transfer coefficient *(U)* at the heat exchanger must be considered for the dynamical system of both the fluids in the component loops. The conjugate wall effect is neglected in the present study.

$$\frac{1}{U} = \frac{1}{h_1} + \frac{1}{h_2} \tag{23}$$

*2.5.1 Laminar regime*

The heat transfer coefficient for a single sided heating used in the present study was taken from the experimental study conducted by Ramakrishnan et al. (2023) for a channel of $\alpha = 0.43$. The correlation shown in (Eq. 24) was obtained from the experimental data in the laminar flow regime for a constant heat flux condition.

$$Nu = 0.012Re + 9.2308 \tag{24}$$

$$Nu = \frac{hD_h}{k} \qquad Re = \frac{u(t)D_h}{\nu} \tag{25}$$

From Eq. (24) it is evident that a change in the Reynolds number ($Re = 200 - 1100$) causes a change in the Nusselt number ($Nu = 11.6 - 22.4$) which indicates that the flow is thermally developing. The *Nu* values (from Eq. 24) are in line with the trends reported by Lee and Garimella (2006) for developing flow. Their study was conducted for three-sided heating; hence it can be observed that the increase in the number of heated sides result in a lower value of the Nusselt number.

*2.5.2 Transition and Turbulent regime*

For both the transition and turbulent regimes, the heat transfer coefficient was derived from the Petukhov correlation using Gnienlinski modification. A correlation for the Nusselt number as a function of friction factor, shown in Eq. (26), was presented by Saha et al. (2015). The friction factors obtained using equations in section 2.4 are used to evaluate the Nusselt number.

$$Nu = \frac{hD_h}{k} = \frac{\left(\frac{f_F}{8}\right)(Re - 1000)Pr}{1 + 12.7\left(\frac{f_F}{8}\right)^{\frac{1}{2}}(Pr^{\frac{2}{3}} - 1)} \qquad (26)$$

*2.6. Bend loss correlations*

The momentum equation incorporates the term for losses occurring due to the bends in the component loops. In our study, we have considered bend losses at 90-degree elbow bends (Threaded and standard) at the four corners ($n = 4$). The bend loss coefficient ($K$) is evaluated based on the 3 K method ($K_1$, $K_\infty$, and $K_d$), which was presented by Darby (2017) and it was concluded that this correlation is valid in all the flow regimes. Based on this reference, the 3 K values are $K_1 = 800$, $K_\infty = 0.14$, and $K_d = 4$ (where, $r/D_h = 1$). The following empirical relation represents the evaluation of bend loss coefficient ($K$):

$$K = \frac{K_1}{Re} + K_\infty \left(1 + \frac{K_d}{D_n^{0.3}}\right) \qquad (27)$$

where, $D_n$ is the hydraulic diameter in inches.

Dass and Gedupudi (2019) acquired experimental data of Bend loss coefficient ($K$) corresponding to different Reynolds number values. These values of $K$ were plotted against Reynolds number and compared with the values of $K$ obtained from 3 K method. The validation concluded that the 3 K method gave a decent estimation of the bend loss coefficient.

*2.7. ODE representation of NCL and CNCL system*

In order to obtain the solutions for momentum and energy equations, the partial differential equations (PDEs) were reduced to a system of stiff ordinary differential equations (ODEs) by replacing the temperature function and piecewise boundary conditions with their Fourier series expansions as they are periodic for a circulation system. The Fourier series expansion is substituted back into the momentum and energy equations.

The functions representing the geometry and coupling are reduced to:

$$\omega_1(x) = \sum_{n=-\infty}^{\infty} \chi_n e^{\frac{in\pi x}{L+L1}} \qquad (28)$$

$$\omega_2(x) = \sum_{n=-\infty}^{\infty} \psi_n e^{\frac{in\pi x}{L+L2}} \tag{29}$$

$$f_1(x) = \sum_{n=-\infty}^{\infty} A_n e^{\frac{in\pi x}{L+L1}} \tag{30}$$

$$f_2(x) = \sum_{n=-\infty}^{\infty} B_n e^{\frac{in\pi x}{L+L2}} \tag{31}$$

The heat flux boundary conditions for NCL and CNCL system are represented as:

*2.7.1 NCL*

*(a) NCL 1 (HHVC):*

$$h_1'(x) = \sum_{n=-\infty}^{\infty} \alpha_n e^{\frac{in\pi x}{L+L1}} \tag{32}$$

$$T_1'(x,t) = \sum_{n=-\infty}^{\infty} \gamma_n(t) e^{\frac{in\pi x}{L+L1}} \tag{33}$$

*(b) NCL 2 (VHHC):*

$$h_2'(x) = \sum_{n=-\infty}^{\infty} \beta_n e^{\frac{in\pi x}{L+L2}} \tag{34}$$

$$T_2'(x,t) = \sum_{n=-\infty}^{\infty} \delta_n(t) e^{\frac{in\pi x}{L+L2}} \tag{35}$$

*2.7.2 CNCL*

*(a) Loop 1:*

$$h_1(x) = \sum_{n=-\infty}^{\infty} \varepsilon_n e^{\frac{in\pi x}{L+L1}} \tag{36}$$

$$T_1(x,t) = \sum_{n=-\infty}^{\infty} \theta_n(t) e^{\frac{in\pi x}{L+L1}} \tag{37}$$

*(b) Loop 2:*

$$h_2(x) = \sum_{n=-\infty}^{\infty} \zeta_n e^{\frac{in\pi x}{L+L2}} \tag{38}$$

$$T_2(x,t) = \sum_{n=-\infty}^{\infty} \varphi_n(t) e^{\frac{in\pi x}{L+L2}} \tag{39}$$

The Fourier series-based expansion uses a spectral method approach which relies on node independence test as opposed to grid independence test. It considers the nodes as basis functions to accurately represent the target function. In the present study, the number of nodes for evaluation purpose is restricted to three to simplify the mathematical model. According to Fichera and Pagano (2003) the first three nodes are sufficient to capture the dynamics of the individual loops, which are 1-D rectangular NCLs. A node independence test was presented by Dass and Gedupudi (2019) to prove that three nodes are sufficient to represent the long-term transients of a coupled system.

The coefficients of the Fourier series expansion can be expressed as:

$$\chi_n = \frac{1}{2(L+L1)} \oint \omega_1(x) \cdot e^{\frac{-in\pi x}{L+L1}} dx \tag{40}$$

All the other coefficients can be expressed in a similar form. Note that the conjugate of the coefficients $\overline{(\theta_n)}$ can be expressed as $\overline{(\theta_n)} = \theta_{-n}$. Each of the coefficient is represented as a complex number (E.g., $\theta_n = x_n + iy_n$). Based on this, the below equations can be split into their real and imaginary parts and solved independently and simultaneously using a mathematical solver.

NCL 1:

$$\frac{du_1'(t)}{dt} + \frac{2P}{D_h}\left(\frac{v}{D_h}\right)^b (u_1')^{2-b} = g\beta \sum_{n=-3}^{n=3} \gamma_n(t) A_{-n} - \frac{nK u_1'^2}{4(L+L1)} \tag{41}$$

$$\frac{d(\gamma_0(t))}{dt} = \alpha_0 \tag{42}$$

$$\frac{d(\gamma_1(t))}{dt} + \frac{i\pi}{L+L1} u_1'(t)\gamma_1(t) + \frac{a\pi^2}{(L+L1)^2}\gamma_1(t) = \alpha_1 \tag{43}$$

$$\frac{d(\gamma_2(t))}{dt} + \frac{2i\pi}{L+L1} u_1'(t)\gamma_2(t) + \frac{4a\pi^2}{(L+L1)^2}\gamma_2(t) = \alpha_2 \tag{44}$$

$$\frac{d(\gamma_3(t))}{dt} + \frac{3i\pi}{L+L1} u_1'(t)\gamma_3(t) + \frac{9a\pi^2}{(L+L1)^2}\gamma_3(t) = \alpha_3 \tag{45}$$

NCL 2:

$$\frac{du_2'(t)}{dt} + \frac{2P}{D_h}\left(\frac{v}{D_h}\right)^b (u_2')^{2-b} = g\beta \sum_{n=-3}^{n=3} \delta_n(t) B_{-n} - \frac{nK u_2'^2}{4(L+L2)} \tag{46}$$

$$\frac{d(\delta_0(t))}{dt} = \beta_0 \tag{47}$$

$$\frac{d(\delta_1(t))}{dt} + \frac{i\pi}{L+L2} u_2'(t)\delta_1(t) + \frac{a\pi^2}{(L+L2)^2}\delta_1(t) = \beta_1 \tag{48}$$

$$\frac{d(\delta_2(t))}{dt} + \frac{2i\pi}{L+L2} u_2'(t)\delta_2(t) + \frac{4a\pi^2}{(L+L2)^2}\delta_2(t) = \beta_2 \tag{49}$$

$$\frac{d(\delta_3(t))}{dt} + \frac{3i\pi}{L+L2} u_2'(t)\delta_3(t) + \frac{9a\pi^2}{(L+L2)^2}\delta_3(t) = \beta_3 \tag{50}$$

CNCL (Loop 1):

$$\frac{du_1(t)}{dt} + \frac{2P}{D_h}\left(\frac{v}{D_h}\right)^b (u_1)^{2-b} = g\beta \sum_{n=-3}^{n=3} \theta_n(t)A_{-n} - \frac{nKu_1^2}{4(L+L1)} \tag{51}$$

$$\frac{d(\theta_0(t))}{dt} = \varepsilon_0 - \sum_{n=-3}^{n=3} \frac{U}{\rho_1 C_{p1} w} \chi_{-n}(\theta_n(t) - \varphi_n(t)) \tag{52}$$

$$\frac{d(\theta_1(t))}{dt} + \frac{i\pi}{L+L1} u_1(t)\theta_1(t) + \frac{a\theta_1(t)\pi^2}{(L+L1)^2} = \varepsilon_1 - \sum_{n=-3}^{n=3} \frac{U}{\rho_1 C_{p1} w} \chi_{1-n}(\theta_n(t) - \varphi_n(t)) \tag{53}$$

$$\frac{d(\theta_2(t))}{dt} + \frac{2i\pi}{L+L1} u_1(t)\theta_2(t) + \frac{4a\theta_2(t)\pi^2}{(L+L1)^2} = \varepsilon_2 - \sum_{n=-3}^{n=3} \frac{U}{\rho_1 C_{p1} w} \chi_{2-n}(\theta_n(t) - \varphi_n(t)) \tag{54}$$

$$\frac{d(\theta_3(t))}{dt} + \frac{3i\pi}{L+L1} u_1(t)\theta_3(t) + \frac{9a\theta_3(t)\pi^2}{(L+L1)^2} = \varepsilon_3 - \sum_{n=-3}^{n=3} \frac{U}{\rho_1 C_{p1} w} \chi_{3-n}(\theta_n(t) - \varphi_n(t)) \tag{55}$$

CNCL (Loop 2):

$$\frac{du_2(t)}{dt} + \frac{2P}{D_h}\left(\frac{v}{D_h}\right)^b (u_2)^{2-b} = g\beta \sum_{n=-3}^{n=3} \varphi_n B_{-n} - \frac{nKu_2^2}{4(L+L2)} \tag{56}$$

$$\frac{d(\varphi_0(t))}{dt} = \zeta_0 + \sum_{n=-3}^{n=3} \frac{U}{\rho_2 C_{p2} w} \psi_{-n}(\theta_n(t) - \varphi_n(t)) \tag{57}$$

$$\frac{d(\varphi_1(t))}{dt} + \frac{i\pi}{L+L2} u_2(t)\varphi_1(t) + \frac{a\varphi_1(t)\pi^2}{(L+L2)^2} = \zeta_1 + \sum_{n=-3}^{n=3} \frac{U}{\rho_2 C_{p2} w} \psi_{1-n}(\theta_n(t) - \varphi_n(t)) \tag{58}$$

$$\frac{d(\varphi_2(t))}{dt} + \frac{2i\pi}{L+L2} u_2(t)\varphi_2(t) + \frac{4a\varphi_2(t)\pi^2}{(L+L2)^2} = \zeta_2 + \sum_{n=-3}^{n=3} \frac{U}{\rho_2 C_{p2} w} \psi_{2-n}(\theta_n(t) - \varphi_n(t)) \tag{59}$$

$$\frac{d(\varphi_3(t))}{dt} + \frac{3i\pi}{L+L2} u_2(t)\varphi_3(t) + \frac{9a\varphi_3(t)\pi^2}{(L+L2)^2} = \zeta_3 + \sum_{n=-3}^{n=3} \frac{U}{\rho_2 C_{p2} w} \psi_{3-n}(\theta_n(t) - \varphi_n(t)) \tag{60}$$

*2.8. Solution*

The reduced system of stiff ODEs obtained from the PDEs mentioned above cannot be solved analytically. A numerical technique was employed to solve the system of equations simultaneously. In our current study, we have utilised SSPRK22 solver from Different Equations package in open-source Julia 1.9.0 to obtain the solutions. The solver stands for Strong Stability Preserving Runge-Kutta (SSPRK) methods for conservation laws, which can be used for ODE systems involving stability analysis for a longer time duration and celestial problems, hence ideal for solving our current problem of dynamical CNCL system. It demonstrates the ability to solve the system of ODEs with a fixed time step ($\Delta t$) which is taken as one second for all the numerical simulations. The fixed time step of $\Delta t = 1s$ provided non-divergent solutions at every iteration.

The temperature distribution in of loop 1 and loop 2 of the CNCL system is given as:

$$T_1(x,t) = \sum_{n=-3}^{3} \theta_n(t) e^{\frac{in\pi x}{L+L1}} \tag{61}$$

$$T_2(x,t) = \sum_{n=-3}^{3} \varphi_n(t) e^{\frac{in\pi x}{L+L2}} \tag{62}$$

$$\begin{aligned}T_1(x,t) = {}& \theta_0(t) \\ &+ 2\left(x_1 \cos\left(\frac{\pi x}{L+L1}\right) + x_2 \cos\left(\frac{2\pi x}{L+L1}\right)\right. \\ &+ x_3 \cos\left(\frac{3\pi x}{L+L1}\right) - y_1 \sin\left(\frac{\pi x}{L+L1}\right) - y_2 \sin\left(\frac{2\pi x}{L+L1}\right) - y_3 \sin\left(\frac{3\pi x}{L+L1}\right)\Bigg)\end{aligned} \tag{63}$$

A similar equation can be obtained for the temperature distribution in loop 2. $\theta_0(t)$ and $\varphi_0(t)$ represent the average temperatures of loop 1 and loop 2 respectively, which change with respect to time.

## 3. Validation

*3.1. NCL*

The SSPRK22 solver implemented using open-source Julia 1.9.0 was validated for NCLs with configuration HHVC (Loop 1) and VHHC (Loop 2) using Vijayan's correlation (Vijayan, 2002). The validation was done for

NCLs with $\alpha = 0.43$, and corresponding width ($w$) and depth ($d$) of 16.125mm and 37.5 mm, respectively. Vijayan's correlation for the Reynolds number ($Re$) and the Modified Grashof number ($Gr_m$) for a steady-state flow is represented as:

$$Re_{ss} = C \left(\frac{Gr_m}{N_g}\right)^r \tag{64}$$

$$C = \left(\frac{2}{P}\right)^r \text{ and } r = \frac{1}{(3-b)} \tag{65}$$

The detailed derivation of this correlation and information on the associated variables ($Gr_m$ and $N_g$) can be found in the corresponding literature. The values of $P$ and $b$ were appropriately taken from Section 2.4 for different flow regimes to incorporate in the correlation. The validation with the correlation justifies the heat flux boundary condition considered for the individual and component loops of the CNCL systems. The dynamic behaviour at steady state is expected to be the same at constant heat flux boundary condition (current study) and constant temperature boundary condition (Vijayan, 2002) at the sink, which verifies that the steady-state behaviour is independent of the boundary condition provided (constant flux or temperature).

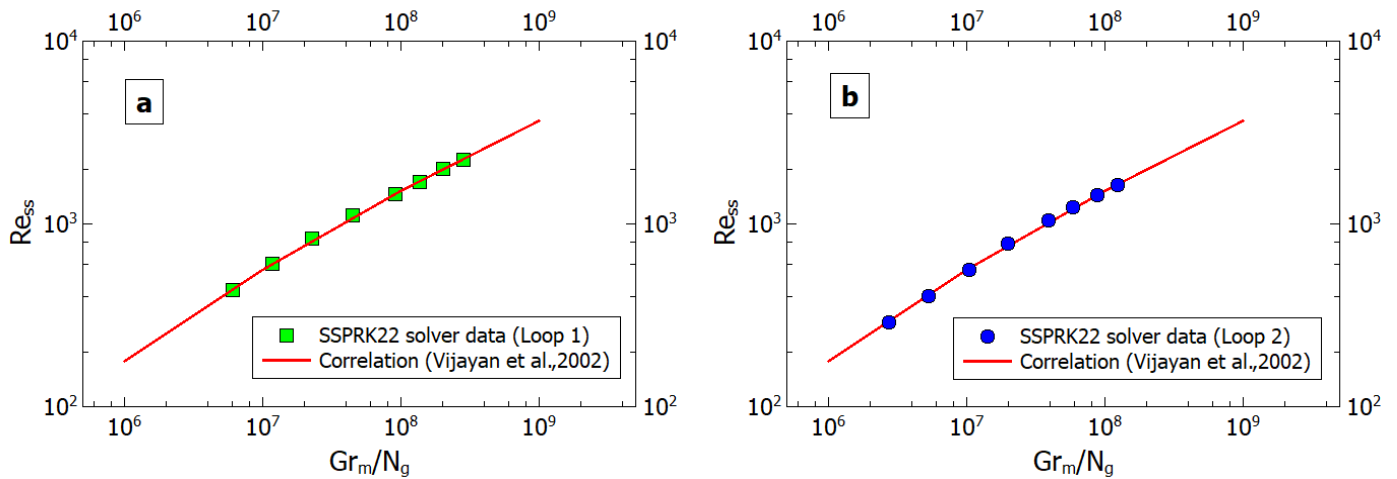

Fig.2. Validation of SSPRK22 solver methodology (current study) with correlation, (Vijayan, 2002) for independent NCL system (a) Loop 1 (HHVC) (b) Loop 2 (VHHC).

*3.2. CNCL*

The CNCL system comprising two independent NCLs (HHVC and VHHC) coupled using a common heat exchanger was validated for steady state condition with the work of Dass and Gedupudi (2019) for laminar regime and incorporating the fluid axial conduction effects and bend loss. Their study considered individual square loops with length ($L1 = L2 = 0.92\ m$) and height ($L$) as 0.92 m and a flow cross-section of 0.04x0.04 (m$^2$). The values of $P$ and $b$ in the friction factor term was taken as 14.23 and 1, respectively, for a square cross-section. The validation for chaotic oscillations was done with the work of Dass and Gedupudi (2022). Fig.3 shows the graphical representation of the validation of velocity at steady state (Fig.3(a)), velocity at chaotic oscillations (Fig.3(b)), and average temperature ratios (Fig.3(c)), of SSPRK22 solver methodology and the CFD study of CNCL. Fig. 3(a) and (c) have been validated with Dass and Gedupudi (2019) where CNCL (a) comprises identical fluid in both the loops and CNCL (b) consists of different fluids in each of the loops. Fig. 3(b) has been validated with Dass and Gedupudi (2022).

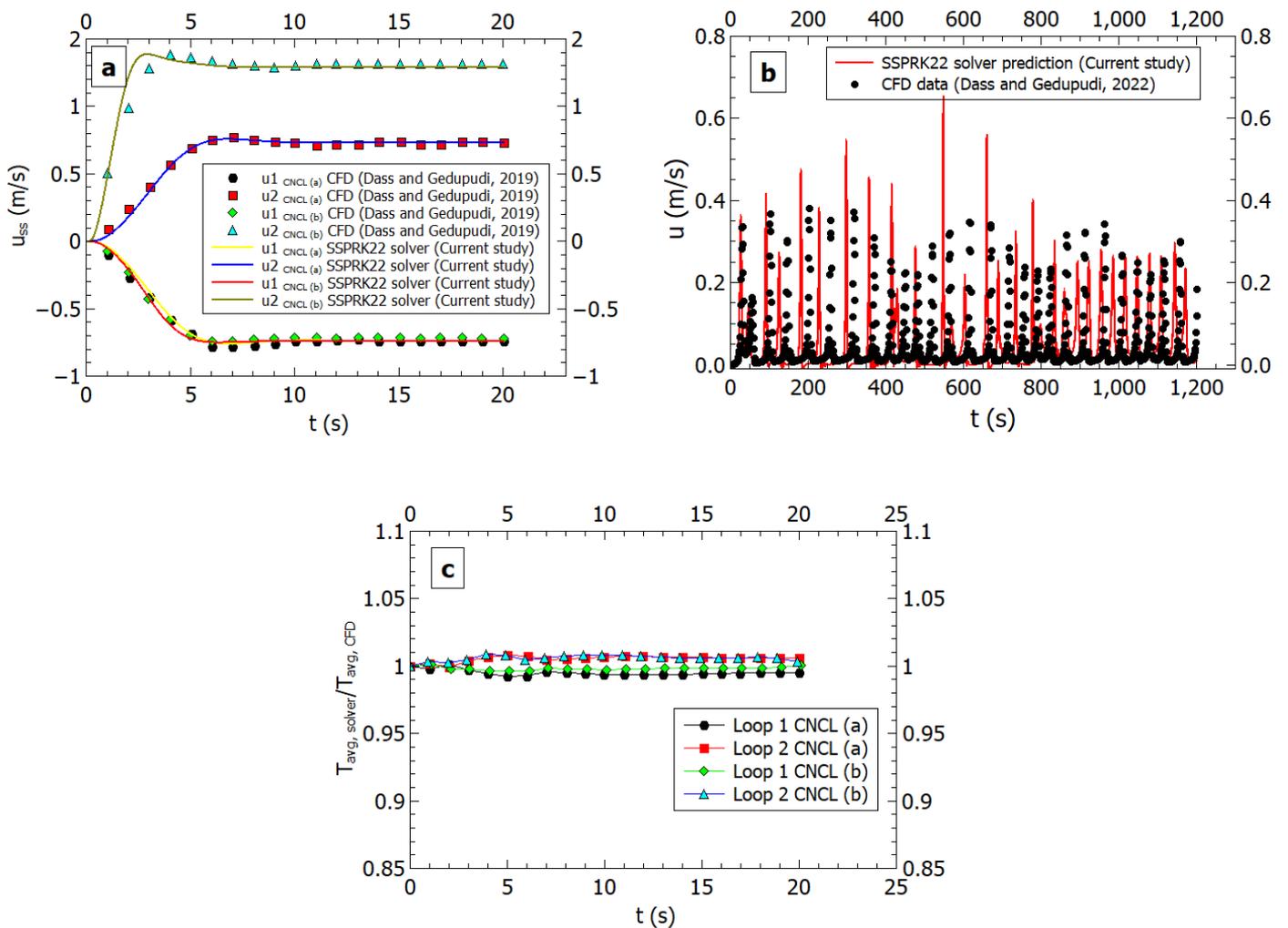

Fig.3. Validation of SSPRK22 solver methodology (current study) with CFD study (Dass and Gedupudi, 2019; Dass and Gedupudi, 2022) for CNCL system (a) Velocity at steady state (b) Velocity for chaotic oscillations (c) Average temperature ratios.

## 4. Results and Discussion

The independent NCLs were coupled using a common heat exchanger and the long-term transient behaviour was observed to study the effect of coupling. In the present study the coupling effect has been discussed by conducting a comprehensive analysis of the buoyancy and viscous forces, heat transfer rate at the heat exchanger section and a comparison with thermally nearly decoupled CNCL system with overall heat transfer coefficient ($U$) = 10 W/m$^2$-K. The buoyancy and viscous forces were evaluated per unit volume of the fluid flowing through the cross section. Long-term transient dynamics of independent and component NCLs were observed for 1 00 000 seconds. Table 1 presents the long-term behaviour of independent and component loops in the coupled system for different cross-sectional areas at a constant heat flux($Q$) = 100 W/m$^2$, aspect ratio ($\alpha$) = 0.43, loop height ($L$) = 0.9 m and length ($L1$ and $L2$) = 1 m. The study has been conducted for a low heat flux value (100 W/m$^2$) to maintain the single-phase flow of water i.e., to avoid temperature fluctuations that lead to two-phase flow. Pressurised water or a liquid metal with sufficient temperature range (where it exists in a single-phase) can also be considered. The thermo-physical fluid properties in component loops are updated at each time step according to the changes occurring in the loop average temperature values.

**Table: 1. Transient behaviour of individual and coupled NCLs.**

| Case | w(m) | d(m) | NCL 1 | NCL 2 | CNCL |
|---|---|---|---|---|---|
| 1 | 0.005375 | 0.0125 | Stable | Stable | Unstable |
| 2 | 0.01075 | 0.025 | Stable | Stable | Neutrally Stable |
| 3 | 0.0215 | 0.05 | Stable | Stable | Stable |
| 4 | 0.03225 | 0.075 | Stable | Stable | Stable |
| 5 | 0.043 | 0.1 | Unstable | Stable | Neutrally Stable |
| 6 | 0.05 | 0.1163 | Unstable | Unstable | Unstable |

*4.1. Case – 1: NCL 1: Stable, NCL 2: Stable, CNCL: Unstable (w = 0.005375 m and d = 0.0125 m)*

Figure 4(a) depicts the long-term stability attained by the individual NCLs. The reason for this stable behaviour can be accredited to the balance established between the driving (buoyancy) and opposing (viscous and bend losses) forces, hence the occurrence of natural circulation at a constant velocity. These forces have been evaluated for per unit volume of the fluid flow. On the contrary, under identical initial and boundary conditions the coupled system comprising the same individual NCLs exhibits instability. Figures 4(b)-(f) provide a holistic view of the instability observed in the component loops of the CNCL system where the corresponding individual NCL systems exhibit a stable behaviour. Figure 4(b) shows the chaotic oscillations of velocity in the component loops due to unbalanced forces, the proof of which can be visualised from Fig. 4(c) and (d). It is observed that the difference between the buoyancy ($f_B$) and viscous

($f_V$) force per unit volume is a finite value throughout the time frame considered for the study. The driving forces are directly dependent on the temperature as can be seen from Eq. (2), and the viscous forces are dependent on the flow velocity. The average temperatures of the component loops (obtained from Eq. 63, Fig. 4(e)) shows fluctuations which represents the occurrence of higher and lower temperatures at certain regions across the loop, indicating the formation of thermal pockets at specific zones which results in the drop of density values and directly affects the buoyancy force term in that region, hence the resulting imbalance in the net force per unit volume term.

The following equations for $f_B$ and $f_V$ are obtained from the conservation of momentum equation (Eq. 2).

$$Buoyancy\ force\ per\ unit\ volume\ (f_B) = \frac{\rho_i g \beta_i}{2(L + L_i)} \left( \oint (T_i - T_{0,i}) f_i(x) dx \right) \tag{66}$$

$$Viscous\ force\ per\ unit\ volume\ (f_V) = \frac{P_{cs} \tau_i}{A_{cs}} + \frac{nK\rho_i u_i^2}{4(L + L_i)} \tag{67}$$

The effect of coupling was also visible at the heat exchanger section. The input heat flux ($Q$) for the individual NCLs and coupled system (at heater section) was kept constant at 100 W/m². At steady state, the heat transfer rate should be equal at the heater (heat input) and heat exchanger section irrespective of the area of heat transfer in the long-term analysis, thus preserving the conservation laws. The heat transfer rate at the heater and the heat exchanger section can be evaluated as:

The heat input at the heater section ($Q_H$):

$$Q_H = Q \times L_h \times 2(w + d) \tag{68}$$

where $L_h$ denotes the heated length, and can be written as: $L_h = L_1 - D_h$

The instantaneous heat transfer at the heat exchanger section ($Q_{HX}$):

$$Q_{HX} = U \times d \times \int_{x=D_h/2}^{x=L-D_h/2} (T_1(x,t) - T_2(x,t))\, dx \tag{69}$$

For this case, where w = 0.005375 m and d = 0.0125 m, the corresponding heat input value was 3.548 W. From Fig. 4(f) it is observed that the heat transfer rate at the heat exchanger section ($Q_{HX}$) is fluctuating and has not attained a steady state. The integral average value for the case shown in Fig. 4(f) is 3.542 W.

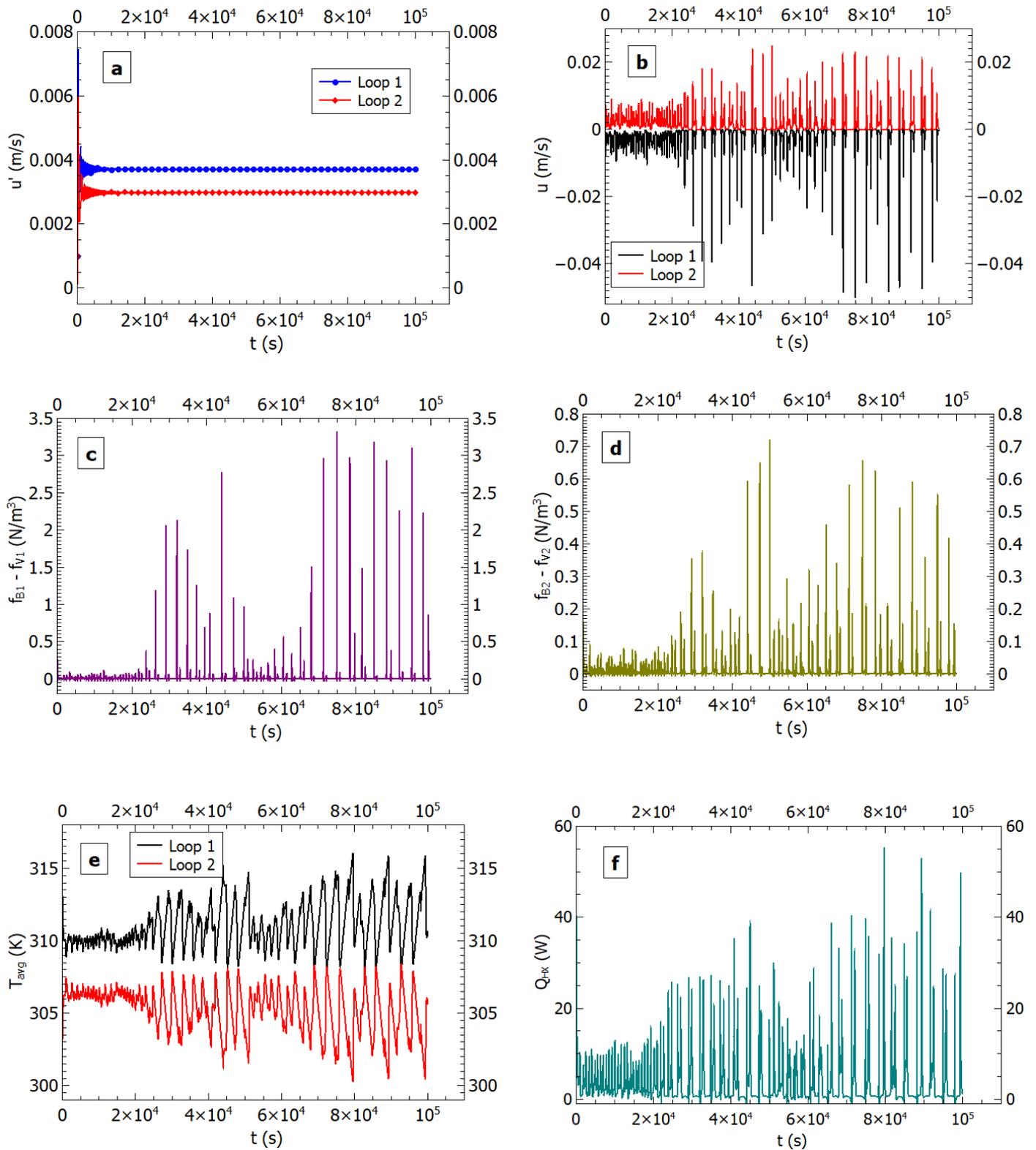

Fig.4. Transient behaviour of NCL and CNCL system for $\alpha = 0.43$, w = 0.005375 m and d = 0.0125 m (a) Velocity of independent loops (b) Velocity of component loops (c) Net force per unit volume (loop 1) (d) Net force per unit volume (loop 2) (e) Average temperature of component loops (f) Heat transfer at the heat exchanger section.

The observations and physics explained using Fig. 4 must be validated and concluded that the chaotic behaviour in the coupled system is arising solely due to the coupling effect. The instability of the component loops is due to the influence of their dynamical behaviour on each other. This is visible from Eq. (8) in Section 2.2, where the heat transfer due to convection ($h_{ci}(x)$) is influenced by the overall heat transfer coefficient ($U$) and average temperature ($T_{i,\,avg}$) difference of the component loops. The overall heat transfer coefficient is used as a tool to control the extent of coupling in the CNCL system due to its physical significance on the heat transfer due to convection. By lowering the value of ($U$) the thermal resistance between the loops is increased and the influence of loop 1 on loop 2 reduces, thus the coupling effect can be controlled. In the existing CNCL system, $U$ is calculated according to Eq. (23), where its value kept changing at every time step due to a change in $h_1$ and $h_2$ (Section 2.5). The developed code was modified to keep the $U$ value at 10 W/m²-K, a much lower value compared to the actual ones ($U$= 381-537 W/m²-K) in the CNCL system. Fig. 5(a) represents the steady state velocity achieved by the component loops which is similar to the one observed in Fig.4(a) of the independent loops. The long-term transients show stable behaviour in both the component loops, and their corresponding net force per unit volume ($f_B - f_V$) Fig. 5(b) and (c) has attained a zero value, and continues to be in that state in the long-term, leading to a constant stable velocity in the loops, a trend that was not observed in Fig. 4(c) and (d). The instantaneous heat transfer at the heat exchanger section was evaluated according to Eq. (69), and it was observed that a constant heat transfer rate of 3.545 W was occurring at the heat exchanger section (Fig. 5(d)) at the steady state. This value shows consistency with the heat input at the heater section (3.548 W) with an error of 0.08%, and it can be inferred that the system has attained a steady state and the heat exchanger section behaves like an imposed boundary condition ($-Q_H$) of the independent NCLs. Therefore, it can be concluded that a lower value of $U$ results in the CNCL system attaining a thermally nearly decoupled state and the component loops have lesser influence on each other's behaviour, and function more as independent NCLs.

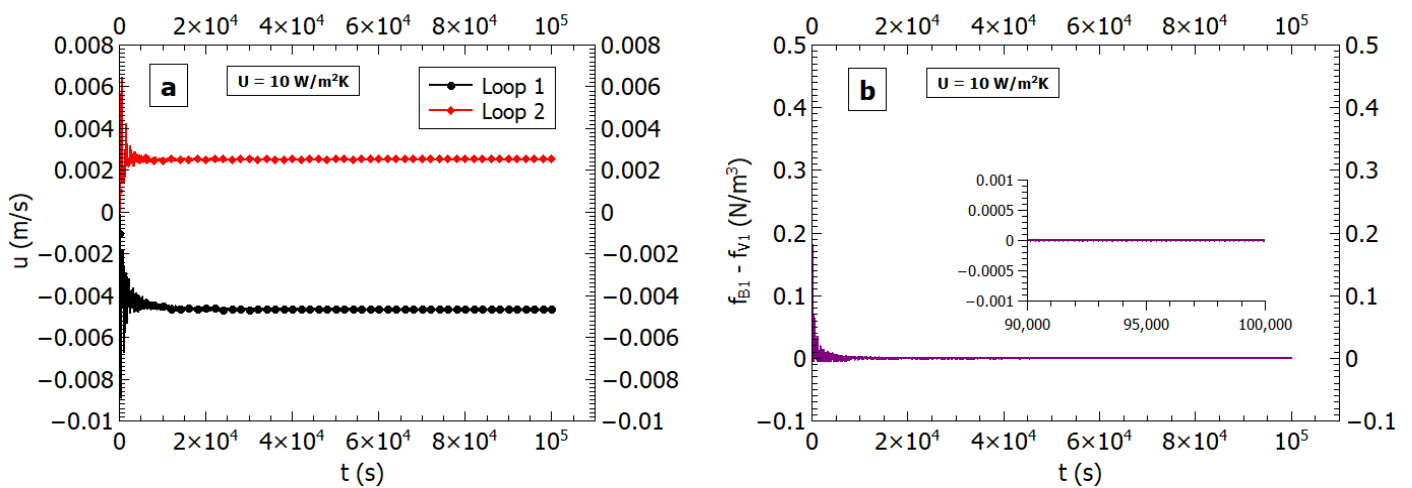

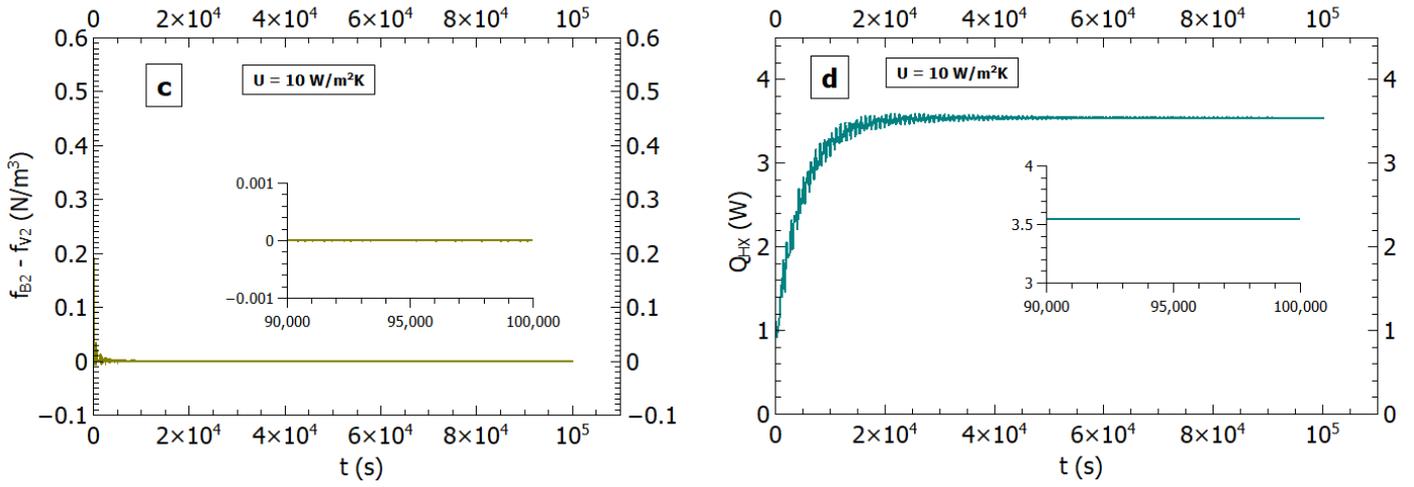

Fig.5. Transient behaviour of CNCL system for $U = 10$ W/m²-K, $\alpha = 0.43$, $w = 0.005375$ m and $d = 0.0125$ m (a) Velocity of component loops (b) Net force per unit volume (loop 1) (c) Net force per unit volume (loop 2) (d) Heat transfer at the heat exchanger section.

*4.2. Case – 3: NCL 1: Stable, NCL 2: Stable, CNCL: Stable (w = 0.0215 m and d = 0.05 m)*

In this case, the cross-sectional area was increased to 0.0215 m x 0.05 m keeping the other parameters constant. Fig. 6(a) shows the graphical representation of the steady state velocity attained by the fluid in loop 1 and loop 2 when they are observed independently. In this configuration, a stable behaviour was established by the coupled system of both the NCLs (Fig. 6(b)) with a constant fluid velocity in the component loops. The physics can be explained similar to the previous case, where the difference between the buoyancy and viscous forces per unit volume of the fluid flow was considered. It is evident from Fig. 6(c) and (d) that the net force has reached a zero value after approximately 30 000 s, and remains in that state for the rest of the time duration (1 00 000 s), hence establishing a constant velocity in the loops. The average temperature plot in Fig. 6(e) represents the principle of energy transport phenomena. The loop 1 initially at a higher temperature (313.15 K) has reached a steady state temperature 309.55 K and the loop 2 (at 303.15 K) has reached a temperature of 306.73 K. At steady state the average temperature is maintained by the component loops as the heat input becomes equal to the heat output, and the overall system behaves as one single NCL. This can be further explained using Fig. 6(f), which illustrates the instantaneous heat transfer rate at the heat exchanger section. The heat input value (Eq. (68)) at the heater ($Q_H$) and the heat rejected at the cooler ($-Q_H$) section is kept constant at 13.87 W. The corresponding heat transfer at the heat exchanger section (Eq. (69)) achieves a value of 13.82 W at steady state, which is considerably close enough with $Q_H$ value with an error of 0.38%, thus validating the conservation law.

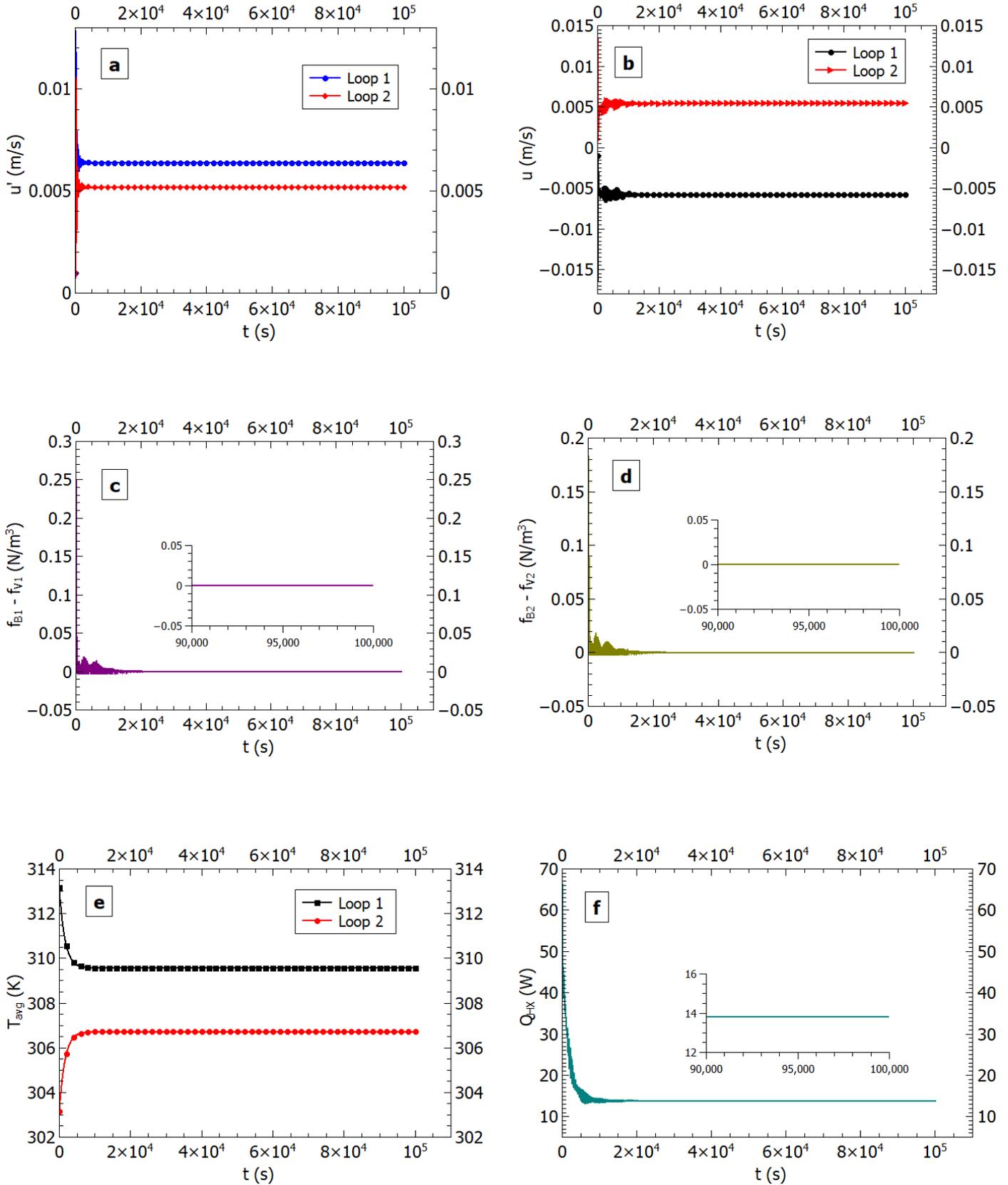

Fig.6. Transient behaviour of NCL and CNCL system for $\alpha = 0.43$, $w = 0.0215$ m and $d = 0.05$ m (a) Velocity of independent loops (b) Velocity of component loops (c) Net force per unit volume (loop 1) (d) Net force per unit volume (loop 2) (e) Average temperature of component loops (f) Heat transfer at the heat exchanger section.

The CNCL system was observed at a lower heat transfer coefficient ($U = 10$ W/m²-K) to analyse the effect of coupling. The component loops in such a state showed significant resemblance to the independent NCL behaviour. This retention of individual behaviour indicates the reduced influence of loop 1 on loop 2 and vice-versa, thus attaining a thermally nearly decoupled state. The fluid flow in the component loops attained a stable velocity Fig. 7(a) after undergoing fluctuations in the initial time frame. The net force per unit volume of the fluid flow in the component loops (Fig. 7(b) and (c)) is identical to the ones from the actual system (Fig. 6(c) and (d)), attaining a zero value at steady state. The energy conservation principle was validated by comparing the $Q_{HX}$ value at steady state for actual system where $U = $ 100-170 W/m²-K and the system forced to a lower heat transfer coefficient ($U = 10$ W/m²-K), with the same heat input ($Q_H$) value at the heater section. The $Q_{HX}$ value for the latter system at steady state is 13.77 W (Fig. 7(d)), which is remarkably close to the $Q_{HX}$ of the actual system (13.82W), and heat input ($Q_H$) at the heater section (13.87 W), with an error of 0.72%.

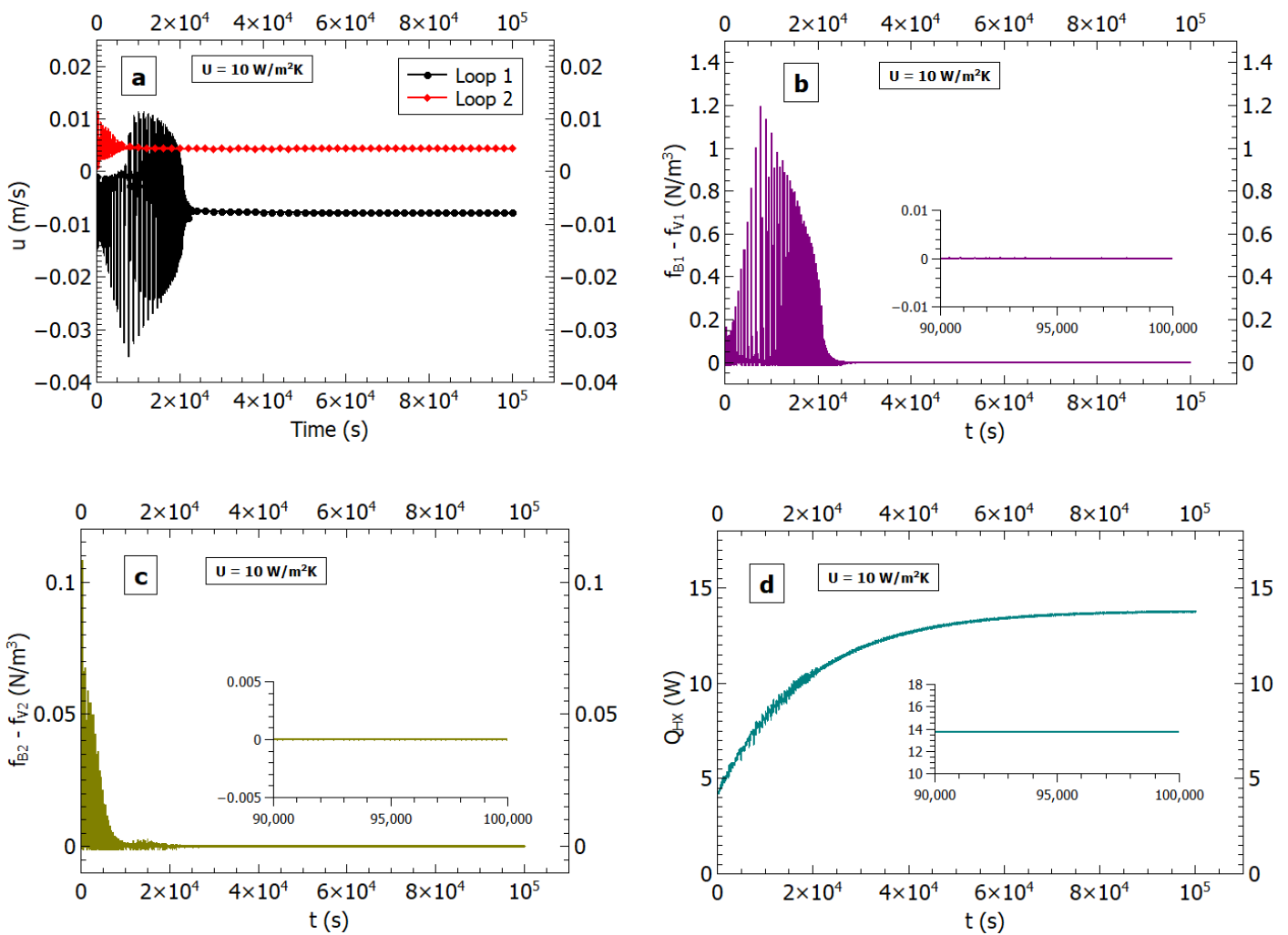

Fig.7. Transient behaviour of CNCL system for $U = 10$ W/m²-K, $\alpha = 0.43$, $w = 0.0215$ m and $d = 0.05$ m (a) Velocity of component loops (b) Net force per unit volume (loop 1) (c) Net force per unit volume (loop 2) (d) Heat transfer at the heat exchanger section.

## 4.3. Case – 5: NCL 1: Unstable, NCL 2: Stable, CNCL: Neutrally stable (w = 0.043 m and d = 0.1 m)

This case was studied for $w = 0.043$ m and $d = 0.1$ m, keeping the other parameters constant. Fig. 8(a) portrays the velocity in the individual loops when observed independently without the coupling effect. It is visible that the fluid flow in loop 1 is chaotic whereas the flow in loop 2 has attained a stable steady state velocity. It can be concluded that for this configuration NCL 1 is unstable and NCL 2 is stable. The velocity in the coupled system exhibits neutrally stable oscillations in both the component loops, which denotes that the instability in loop 1 has been dampened due to the influence of thermal coupling, i.e., effect of component loops on each other, and is evident from Fig. 8(b). The component loops are influenced by the imbalance of the net forces, which is directly affected by the temperature fluctuations and flow velocity (Section 4.1). The finite value of the net force per unit volume (Fig. 8(c) and (d)) with a constant amplitude shows that the fluid flow experiences unbalanced forces of fixed magnitudes. To attain a steady-state the momentum equation (Eq. 2) should be balanced to make the $\frac{du_i(t)}{dt}$ term zero, which implies that the value of $f_B - f_V$ should tend to zero (Section 4.1). Since this was not achieved by the system it can be inferred that the neutrally stable oscillations in the component loops are a direct result of a finite value net force per unit volume term with constant amplitude fluctuations in the momentum equation.

The average temperature plot (Fig. 8(e)) shows that both the loops of the coupled system has attained a steady-state. The local temperatures attained by the component loops have a direct impact on each other's behaviour (Section 4.1). The temperature dependent thermophysical properties of the fluid influences the heat transfer due to convection term, which is also affected by $U$. All the parameters in the $h_{ci}(x)$ equation are temperature dependent (except heat transfer area), and are responsible for the dynamical behaviour of the coupled system. The overall behaviour of the system can be attributed to these factors.

Fig. 8(f) shows the temporal distribution of the instantaneous heat transfer at the exchanger section. The heat input was evaluated as 26.88 W (Eq. 68) at the heater section. The values of $Q_{HX}$ from the plot exhibits a neutrally stable trend, with a constant amplitude of 0.767 W (Max. = 27.05 W and Min. = 26.29 W). This trend can be attributed to the fluctuating $U$ values. The time integral average value of $Q_{HX}$ over the domain of neutrally stable behaviour in Fig. 8(f) (t (s) = 98 000 – 1 00 000) is 26.72 W, which is close to the heat input (26.88 W) with an error of 0.59%.

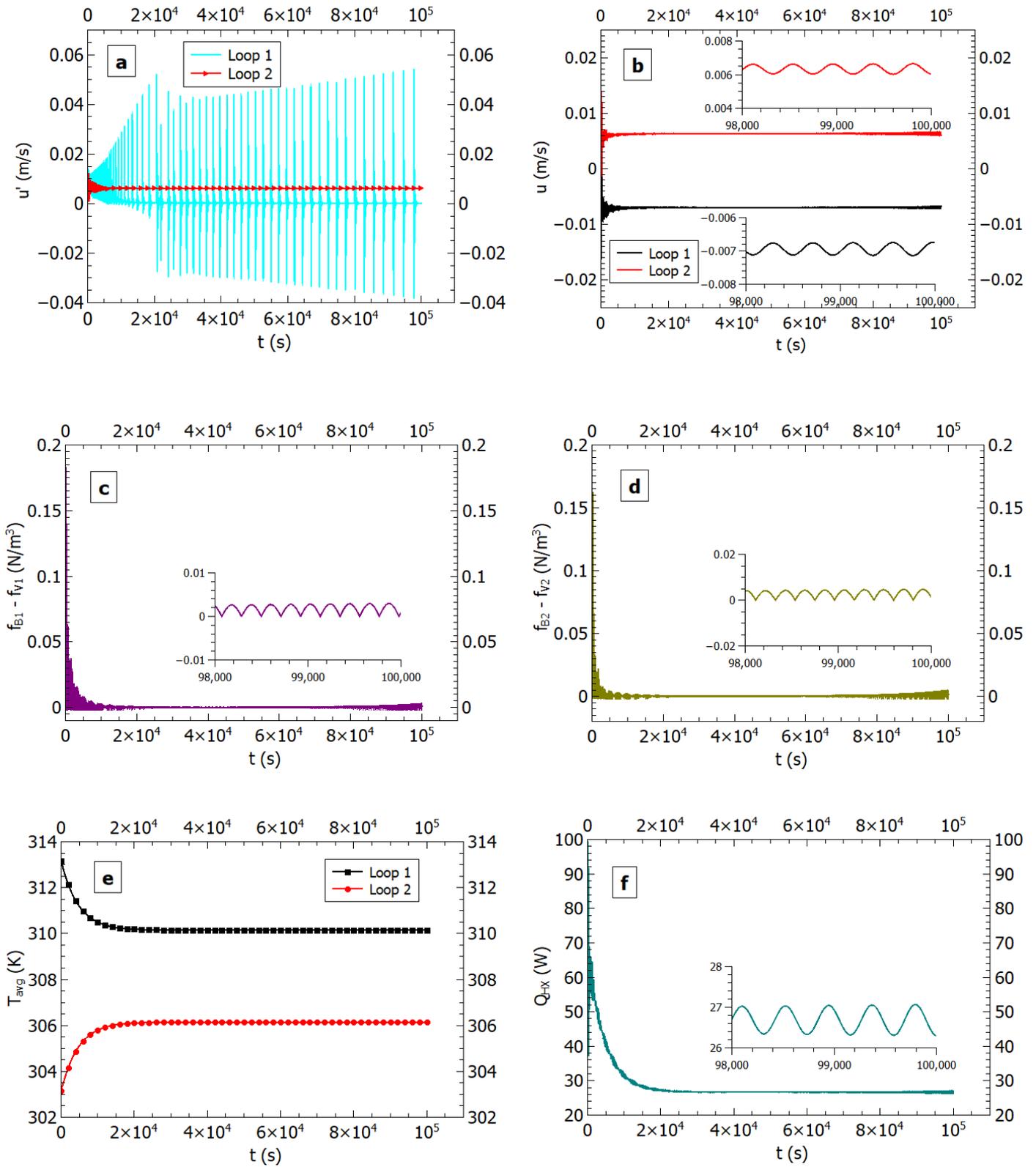

Fig.8. Transient behaviour of NCL and CNCL system for $\alpha = 0.43$, $w = 0.043$ m and $d = 0.1$ m (a) Velocity of independent loops (b) Velocity of component loops (c) Net force per unit volume (loop 1) (d) Net force per unit volume (loop 2) (e) Average temperature of component loops (f) Heat transfer at the heat exchanger section.

In order to verify the physics of the CNCL system a lower $U$ ($U$ = 10 W/m²-K) was imposed as opposed to the actual system's values ($U$ = 48-123 W/m²-K) to verify the retention of independent NCL behaviour of the component loops. Fig. 9(a) shows that the velocity of component loop 1 is chaotic whereas loop 2 has achieved a stable steady-state, similar to the plot that was observed in Fig. 8(a) for independent NCLs.

The corresponding net force per unit volume plot for loop 1 (Fig. 9(b)) has irregular fluctuations (0- 3 N/m³) and accounts for the dynamical instability of fluid flow. For loop 2, (Fig. 9(c)) the initial fluctuations decrease over time and eventually attain a zero value after ~55 000 s, hence the corresponding flow velocity attains a steady state. This can be easily verified by looking at Fig. 9(a) and (c) for comparison with respect to time at which steady state was achieved.

Fig. 9(d) shows the $Q_{HX}$ value calculated at each time step. The value increases steadily with constant fluctuations and eventually shows a trend where it oscillates with a constant maximum and minimum value, though the oscillations cannot be termed as neutrally stable. This nature of the graph can be attributed to the contribution from loop 1 which has retained its characteristics of instability. Even though the $Q_{HX}$ value does not attain a neutrally stable state, it can be visually inferred from the inset plot that the mean value tends to ~25.3 W, which is close to the time integral average value of the actual system (26.72 W) and heat input ($Q_H$ = 26.88 W) at the heater section, with an error of 5.3% and 5.8% respectively.

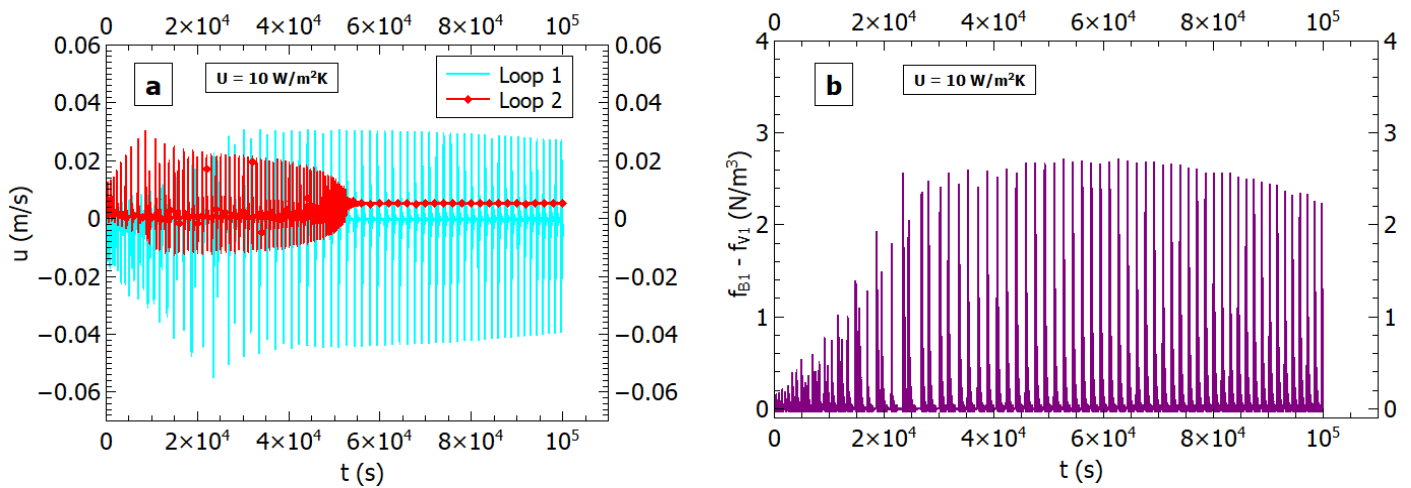

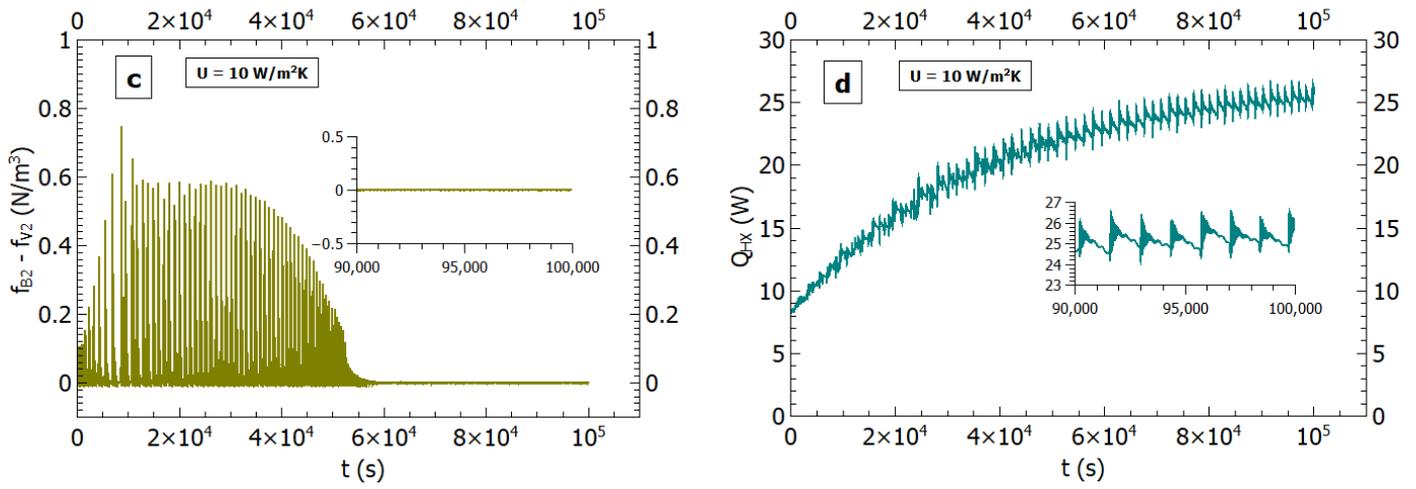

Fig.9. Transient behaviour of CNCL system for $U = 10$ W/m$^2$-K, $\alpha = 0.43$, $w = 0.043$ m and $d = 0.1$ m (a) Velocity of component loops (b) Net force per unit volume (loop 1) (c) Net force per unit volume (loop 2) (d) Heat transfer at the heat exchanger section.

### 4.4. Transient behaviour of other cases

The analysis of long-term transient dynamics of other cases (Table 1) at different flow cross-section areas were conducted in a similar way (Section 4.1-4.3). The system with $w = 0.01075$ m and $d = 0.025$ m exhibited a stable behaviour for both of its individual NCLs, whereas their coupled system revealed neutrally stable oscillations. This trend is quite different from the case observed in Section 4.3, where the independent NCL 1 was unstable and NCL 2 was stable, though their coupled system behaved in a neutrally stable state.

Another case with $w = 0.03225$ m and $d = 0.075$ m displayed stable steady-state behaviour for independent NCLs and component loops in a coupled system, similar to the behaviour observed in Section 4.2. When the cross-section area was increased further to 0.05 x 0.1163 (m$^2$), both NCL 1, NCL 2 along with their coupled system demonstrated chaotic oscillations. The observed behaviour of coupled NCLs for these cases can be explained using the same analysis as presented in Sections 4.1 - 4.3.

*4.5. Effect of thermophysical fluid properties*

The long-term transient dynamics of component loops with constant fluid properties at average fluid temperature ($T_{avg}$ of Loop 1 = 313.15 K and Loop 2 = 303.15 K) was observed for 1 00 000 s to analyse the influence of temporal change in thermophysical fluid properties with respect to $T_{avg}$ in the component loops. For the individual independent NCLs a temporal change in $T_{avg}$ gives a zero value (Eq. 42 and 47), which indicates that the average temperatures of independent NCL 1 and NCL 2 remain constant, hence the corresponding fluid properties and the overall behaviour of the individual NCLs remains unchanged. From Table 2, it can be observed that Cases 1,2 and 5 show a change in the terminal state of CNCL, in terms of the system's behaviour as compared to the ones with variable fluid properties (Table 1). The other cases (3,4 and 6) retain their behaviour, like the one portrayed in Table 1. However, there can be a change in the magnitude (steady-state) and range (neutrally stable or chaotic oscillations) for velocity and temperature profiles.

Table: 2. Transient behaviour of individual and coupled NCLs (Fluid properties at constant $T_{avg}$).

| Case | w(m) | d(m) | NCL 1 | NCL 2 | CNCL |
|---|---|---|---|---|---|
| 1 | 0.005375 | 0.0125 | Stable | Stable | Neutrally Stable |
| 2 | 0.01075 | 0.025 | Stable | Stable | Stable |
| 3 | 0.0215 | 0.05 | Stable | Stable | Stable |
| 4 | 0.03225 | 0.075 | Stable | Stable | Stable |
| 5 | 0.043 | 0.1 | Unstable | Stable | Stable |
| 6 | 0.05 | 0.1163 | Unstable | Unstable | Unstable |

*4.5.1. Case – 1: NCL 1: Stable, NCL 2: Stable, CNCL: Neutrally Stable (w = 0.005375 m and d = 0.0125 m)*

The effect of temperature dependent fluid properties on the long-term transient dynamics of the coupled system is shown in Fig. 10. The component loops with constant fluid properties ($T_{avg,\ loop\ 1}$ = 313.15 K and $T_{avg,\ loop\ 2}$ = 303.15 K) exhibited neutrally stable behaviour (Fig. 10(a)), where the actual system which incorporated temperature dependent fluid properties demonstrated chaotic oscillations. The neutrally stable velocity profile achieved a mean velocity of -0.003 (loop 1) and 0.0022 (loop 2) and has a fixed amplitude with a constant maxima and minima which can be explained using the net force per unit volume term. Fig. 10(b) depicts the local temperature variation at the inlet (c) and exit (d) of the heat exchanger section in loop 1 (Fig. 1(b)) with constant and varying fluid thermophysical properties. It can be observed that the actual system shows unstable temperature fluctuations, which indicates the formation of thermal pockets (Section 4.1). On the other hand, the system at constant fluid properties shows neutrally stable oscillations for the temperature profile at locations 'c' (inlet of the heat exchanger for loop 1) and 'd' (exit of the heat exchanger for loop 1). Fig. 10(c) shows a similar behaviour for loop 2 at 'e' (inlet of the heat exchanger for loop 2) and 'f' (exit of the heat exchanger for loop 2). Therefore, the terminal state of a CNCL system is sensitive to the change in the fluid thermophysical properties.

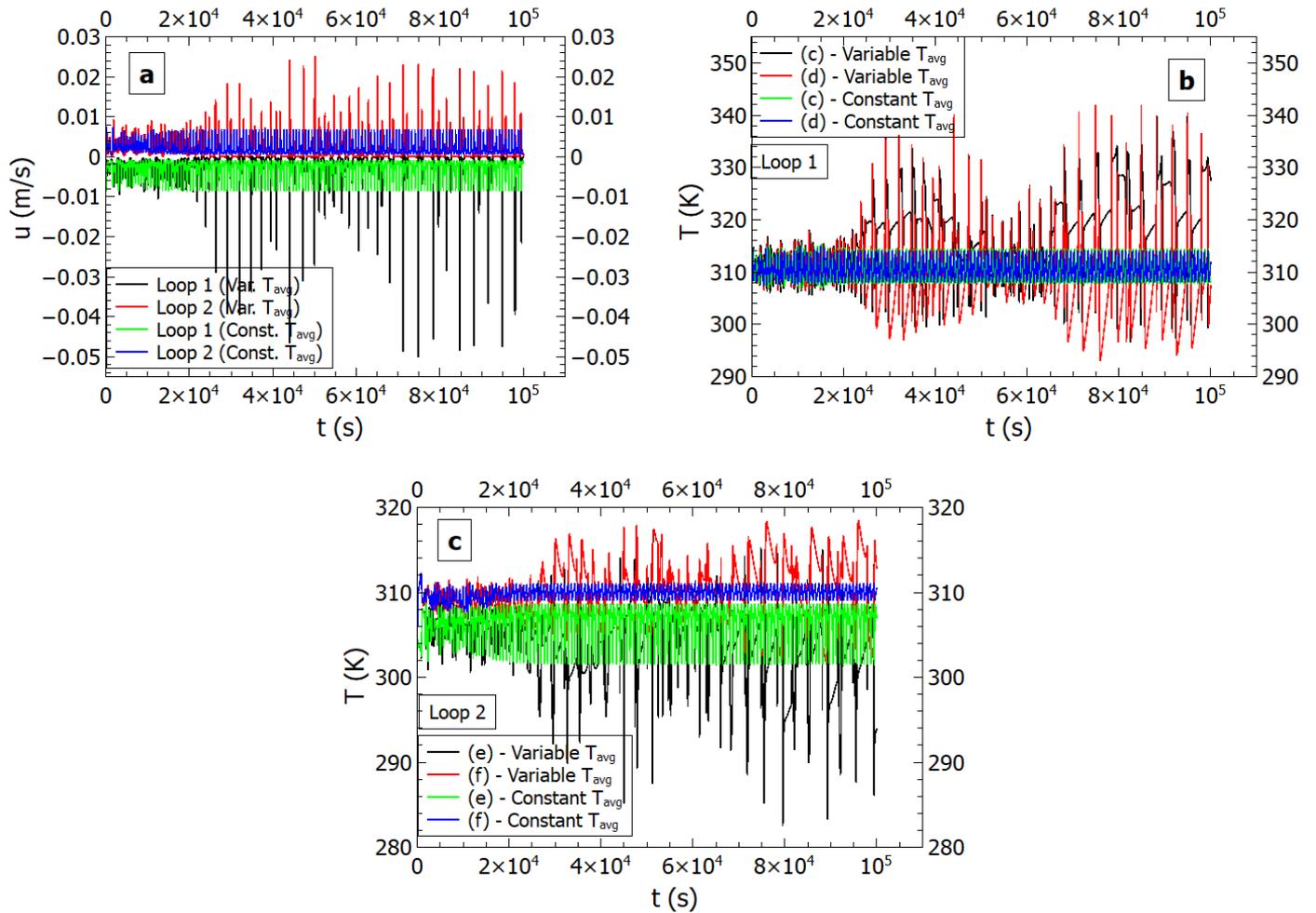

Fig.10. Transient behaviour of CNCL system for $w = 0.005375$ m and $d = 0.0125$ m with constant and varying fluid properties (a) Velocity of component loops (b) Temperature profile at inlet (c) and exit (d) of heat exchanger section (loop 1) (c) Temperature profile at inlet (e) and exit (f) of heat exchanger section (loop 2).

*4.5.2 Case – 3: NCL 1: Stable, NCL 2: Stable, CNCL: Stable ($w = 0.0215$ m and $d = 0.05$ m)*

Based on Fig. 11, the coupled system reaches a stable state with constant temperature fluid properties, resembling the behaviour observed with variable fluid properties (Section 4.2). The velocity in loop 1 is -0.0058 m/s with variable fluid properties and -0.00636 m/s with constant properties. The deviation of the system with constant fluid properties from the actual system with variable fluid properties is approximately 9.81%. Similarly, for loop 2, the steady state velocities with variable and constant fluid properties are 0.0054 m/s and 0.0052 m/s, respectively. The local temperature profile at the inlet and exit of the heat exchanger section for loop 1 and loop 2 is depicted in Fig. 11(b) and (c). The deviation in temperatures is minimal at steady state, with the inlet of loop 1 (location (c) from Fig. 1(b)) at 309.71 K and 309.67 K, and exit (location (d) from Fig. 1(b)) at 309.67 K and 309.65 K with variable and constant fluid properties respectively. A similar observation can be made for loop 2 from Fig. 11(c). Thus, there is no change in the overall system's behaviour with constant and varying fluid properties, but only the magnitudes of velocities and temperatures attained at steady state.

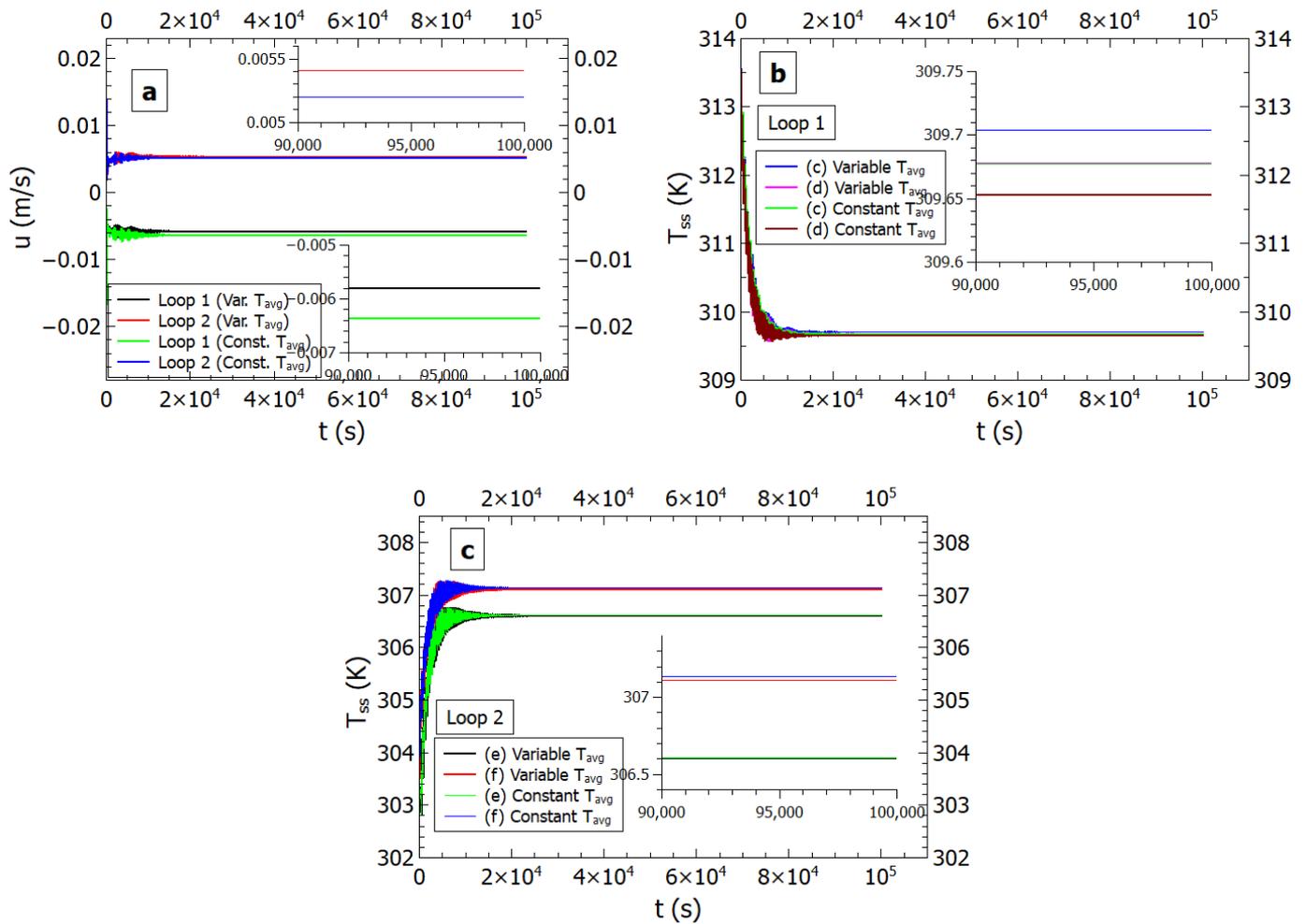

Fig.11. Transient behaviour of CNCL system for $w = 0.0215$ m and $d = 0.05$ m with constant and varying fluid properties (a) Velocity of component loops (b) Temperature profile at inlet (c) and exit (d) of heat exchanger section (loop 1) (c) Temperature profile at inlet (e) and exit (f) of heat exchanger section (loop 2).

*4.5.3 Case – 5: NCL 1: Unstable, NCL 2: Stable, CNCL: Stable (w = 0.043 m and d = 0.1 m)*

This case presents a change in the final state achieved by the component loops in the CNCL system, going from neutrally stable condition (Table 1 and Section 4.3) with variable fluid properties to stable state with constant fluid properties. From Fig, 12(a), the steady state velocity of loop 1 with variable properties oscillates between a constant maximum value of -0.0071 m/s and minimum value of -0.0067 m/s, with a mean of -0.0069 m/s; whereas with constant properties, it achieves a steady-state value of -0.0075 m/s, with an error of ~7.74%. A similar observation and analysis can be done for loop 2 as well. The temperature profiles of loop 1 (Fig. 12(b)) at the inlet (location (c) in Fig. 1(b)) and exit (location (c) in Fig. 1(b)) show a similar behaviour as observed in Fig. 12(a) with neutrally stable oscillations. A small amplitude oscillation is depicted at inlet and exit, with the inlet of loop 1 attaining a maximum of 310.21 K and minimum of 310.2 K. On the other hand, the steady state temperature profile (with constant fluid properties) at the inlet of loop 1 reaches 310.17 K. The deviation of this temperature is only around 0.035 K with the mean value for the variable fluid properties case. The end state of loop 2 exhibits a nature similar to loop 1, and a similar explanation can be offered as above for velocity and temperature profiles (Fig. 12(c)).

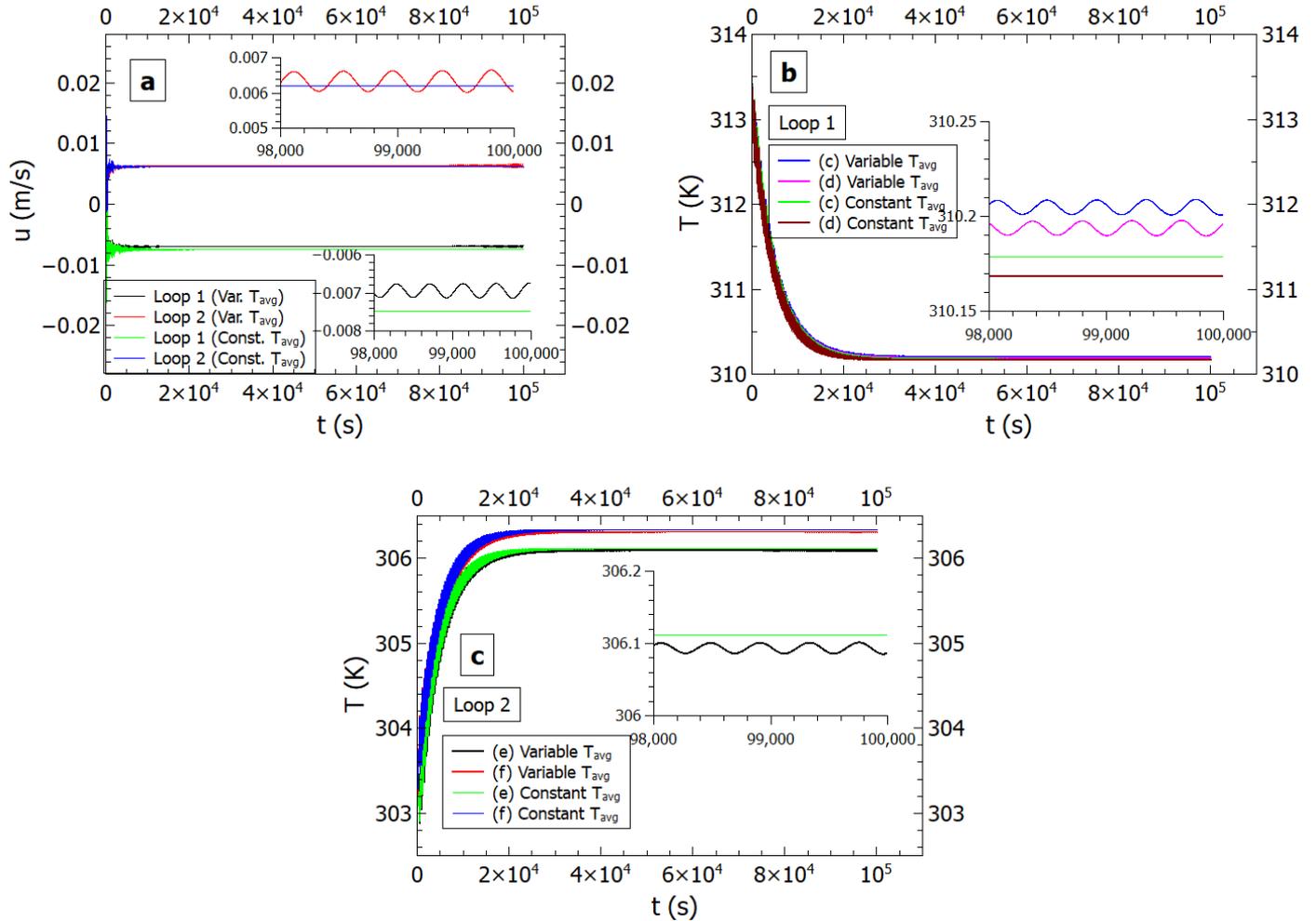

Fig.12. Transient behaviour of CNCL system for $w = 0.043$ m and $d = 0.1$ m with constant and varying fluid properties (a) Velocity of component loops (b) Temperature profile at inlet (c) and exit (d) of heat exchanger section (loop 1) (c) Temperature profile at inlet (e) and exit (f) of heat exchanger section (loop 2).

## 5. Conclusion

The current work presents the effect of coupling on the long-term transient dynamics of the component NCLs of a CNCL system. The CNCL system consists of a flat plate heat exchanger between the component rectangular NCLs of HHVC and VHHC configurations having rectangular cross-section. The analysis was done by modelling the dynamical independent NCLs and its corresponding CNCL system employing a Fourier series based 1-D model. The model incorporates temperature dependent thermophysical properties of the fluid throughout the time frame. The developed model has been verified against the existing relevant literature. The conclusions that can be drawn from the present study are as follows:

1. It can be observed that the 1-D modelled NCL and CNCL system is sufficient to capture the dynamics of the complex system. The developed model using open-source Julia 1.9.0 is highly efficient and illustrates the ability to capture and analyse the long-term dynamics (1 00 000 s), providing a comprehensive understanding of the evolving patterns, thereby eliminating the need of any commercial softwares or time consuming CFD counterparts. The model is significant when it comes to practical applicability as it uses real time working fluid (water) and considers appropriate correlations and preserves the temperature dependent properties over time.

2. The present study reveals the effect of coupling of component loops for different cross-sectional areas at constant heat input and loop dimensions. It is demonstrated that the dynamic response of the coupled system is unpredictable with respect to the individual NCLs' behaviour even when the operating conditions (heat input, initial and boundary conditions) and loop dimensions (loop height and width) are kept constant for a particular flow area.

3. The transient dynamics of the stable and unstable nature was explained using the net force per unit volume ($f_B - f_V$) of the fluid flow. It was shown that this term achieves a zero value to attain a stable steady state, and any finite value can lead to an unstable (irregular fluctuations) or neutrally stable (constant amplitude fluctuations) behaviour. The fluctuations occurring in the buoyancy force per unit volume term was attributed to the change in average temperature values over time.

4. The instantaneous heat transfer occurring at the heat exchanger section was studied in detail for the cases where stability was achieved, to explain the steady state nature of the component loops by comparing with the constant heat input value. It was proved that the extent of coupling and the influence of component loops on each other in a coupled system can be controlled by varying the overall heat transfer coefficient ($U$). The verification of the fact that the terminal state of the CNCL system solely arises from the coupling effect was validated by showing that the component loops nearly retained their independent behaviour, as in a decoupled state, at a lower value of $U$. That is, the influence of loop 1 on loop 2 and vice versa was negligible at a lower value of $U$.

5. For the range of the parameters considered in the study, it can be concluded that the dynamics of the coupled system is also influenced by the fluid properties, and that a stable system remains as a stable system upon changing the fluid properties from variable (temperature dependent) to constant. A change in properties from constant to variable (temperature dependent) tends to make the coupled system unstable, but need not necessarily lead to instability or more instability. It depends on the amplitude of fluctuations and the sensitivity of fluid properties to temperature.

6. The limitations of the current work are that it does not take into account the conjugate wall effects and the loop inclination. The present model can also be used to investigate the effect of inclination by modifying the geometry function $\{f_i(x)\}$. The future work includes incorporating these features into the model.